# Federal Reserve Communication and the COVID-19 Pandemic[*]

Jonathan Benchimol,[1] Sophia Kazinnik[2] and Yossi Saadon[3]

September 2025

## Abstract

In this study, we examine the Federal Reserve's communication strategies during the COVID-19 pandemic, comparing them with communication during previous periods of economic stress. Using specialized dictionaries tailored to COVID-19, unconventional monetary policy (UMP), and financial stability, combined with sentiment analysis and topic modeling techniques, we identify a distinct focus in Fed communication during the pandemic on financial stability, market volatility, social welfare, and UMP, characterized by notable contextual uncertainty. Through comparative analysis, we juxtapose the Fed's communication during the COVID-19 crisis with its responses during the dot-com and global financial crises, examining content, sentiment, and timing dimensions. Our findings reveal that Fed communication and policy actions were more reactive to the COVID-19 crisis than to previous crises. Additionally, declining sentiment related to financial stability in interest rate announcements and minutes anticipated subsequent accommodative monetary policy decisions. We further document that communicating about UMP has become the "new normal" for the Fed's Federal Open Market Committee meeting minutes and Chairman's speeches since the Global Financial Crisis, reflecting an institutional adaptation in communication strategy following periods of economic distress. These findings contribute to our understanding of how central bank communication evolves during crises and how communication strategies adapt to exceptional economic circumstances.

*Keywords:* Central bank communication, Unconventional monetary policy, Financial stability, Text mining, COVID-19.

*JEL Classification:* C55, E44, E58, E63.

[*] This paper does not necessarily reflect the views of the Bank of Israel, the Federal Reserve Bank of Richmond, or the Federal Reserve System. We thank Itamar Caspi (discussant) and participants at the 133rd American Economic Association (AEA/ASSA) annual meeting, the 2021 Central Bank Research Association (CEBRA) annual meeting, and the Bank of Israel research seminars for their useful comments. An extended version of this paper was published in Covid Economics, CEPR Press, issue 79, pages 218-256, May 2021. The replication files for the methodology used in this paper are available at doi.org/10.24433/CO.1132001.v1; Text data are available at github.com/JBenchimol/central-bank-texts

[1] Bank of Israel, Research Department, Jerusalem, Israel. Corresponding author.
  Email: jonathan.benchimol@boi.org.il
[2] Stanford University, Digital Economy Lab, Stanford, CA, USA.
  Email: kazinnik@stanford.edu
[3] Bank of Israel, Research Department, Jerusalem, Israel.
  Email: yosis@boi.org.il

# 1. Introduction

In the wake of the COVID-19 pandemic, central banks globally committed to undertaking "all necessary steps" to alleviate its economic impact, including lowering interest rates to the zero lower bound (Carney, 2020). As the crisis unfolded in March 2020, the Federal Reserve (Fed) and numerous other central banks implemented robust unconventional monetary policy (UMP) measures to surmount the constraints of conventional monetary policy. Communication emerged as a pivotal tool through which these policies were conveyed.

Central banks engage in communication across various topics through distinct channels, guided by well-defined objectives (Hansen et al., 2019; Benchimol et al., 2020). Their communication aims to inform (e.g., current and future policy objectives and decisions), explain (e.g., past, present, and future economic outlooks and decisions), and influence (e.g., current and future uncertainty and financial decisions) economic agents. Typically disseminated in textual form (Haldane and McMahon, 2018), these instances of communication underwent significant adaptations during the COVID-19 pandemic, a period that affected all sectors of the global economy (Chetty et al., 2024).

This descriptive study delves into the distinct communication strategies employed by the Fed during the COVID-19 pandemic compared to past crises. We conduct a comparative analysis of communication instances spanning the past two decades, focusing on three critical periods: the dot-com crisis, the Global Financial Crisis (GFC), and the COVID-19 pandemic. We use text analysis techniques to examine the content, sentiment, and timing of the Fed's communication across various channels, including interest rate announcements, minutes, and speeches.

Our main contribution lies in revealing the more reactive nature of the Fed's communication and actions during the COVID-19 crisis relative to other crises. We uncover significant differences in content, sentiment, and timing, particularly regarding financial stability, market volatility, and UMP. The analysis gauges the extent and intensity of these differences between and during crises. Additionally, we assess the impact of Fed communication on equity volatility in the United States (US), and explore the connection between (conventional) monetary policy and financial stability sentiment (Born et al., 2014; Correa et al., 2021).

By offering a nuanced comparison and novel insights into the Fed's crisis communication strategies, this paper aims to inform future central bank practices during periods of economic turmoil. We do not claim causality between communication and economic outcomes, but we shed light on the Fed's communication patterns and their potential role in shaping expectations about policy responses. Our findings add to the growing literature on central bank communication during crises, which emphasizes the importance of timely and transparent communication in managing market expectations (Blinder et al., 2008; Hansen and McMahon, 2016; Gardner et al., 2022).

Given the zero lower bound (ZLB) of the nominal interest rate, the Fed, like many central banks, resorted to alternative channels, including communication, quantitative easing (QE), balance sheet policies, lending facilities, forward guidance, and other market operations and monetary measures (Bianchi et al., 2020; Guerrieri et al., 2022). Our analysis reveals that Fed communications during the COVID-19 crisis were characterized by uncertainty and heterogeneity, both temporally and across communication types. We examine the efficacy of the Fed's transparent communication in supporting UMP measures amid the challenges posed by the COVID-19 pandemic (Daly, 2020; Craig et al., 2021). Our results suggest that the Fed adeptly employed communication during the COVID-19 crisis, reflecting an increased proficiency in crisis-specific communication management.

Effective central bank communication necessitates clear and timely updates on current and near-term policy actions in an environment fraught with economic uncertainty. Our findings document the prevalence of positive financial stability updates in monetary policy and financial market-related communications after the GFC, and their relatively less negative tone during the COVID-19 crisis, attesting to this shift.[4]

Furthermore, this paper explores the nuanced differences in content, timing, and sentiment of the Fed's communications across crises, suggesting potential implications for financial system developments (Nyman et al., 2021). Since the GFC, communication on UMP has become a "new normal," evident in major communication types such as Federal Funds Rate (FFR) announcements, Federal Open Market Committee (FOMC) minutes, and Fed Chairman speeches. The study also connects conventional monetary policy and financial stability sentiment (FSS).

The subsequent sections of the paper are organized as follows. Section 2 details the data and methodology, Section 3 presents sentiment analysis and topic modeling results, Section 4 examines the Fed's communications regarding UMP, Section 5 discusses the Fed's early communications regarding the pandemic, Section 6 compares the Fed's conventional monetary policy to the FSS over the past two decades, Section 7 discusses policy implications, and Section 8 concludes.

## 2. Data and Methodology
### 2.1 Data
Our investigation centers on primary communications disseminated by the Fed to the public, encompassing instances detailing monetary policy discussions (FFR announcements and FOMC minutes) and speeches by the Fed Chairman from 2000 to 2020. The particulars of our dataset are concisely presented in Table 1.

To augment our analysis, we incorporate data pertinent to the pandemic—specifically, the daily count of new COVID-19 cases—sourced from the COVID-19 Data Repository maintained by the Johns Hopkins University Center for Systems Science and

---

[4] To proxy for the degree of financial stability conveyed in a central bank communication, we calculate a financial stability score for each relevant communication based on a word count of the terms that can also be found in the financial stability dictionary (Correa et al., 2021).

Engineering (CSSE).[5] In addition, we incorporate daily market-based indicators, including the S&P 500 equity index, the Chicago Board of Options Exchange Volatility Index (VIX), the broad nominal effective exchange rate (NEER), and the nominal interest rate (FFR), retrieved from Bloomberg.

**Table 1.** Descriptive Statistics: Federal Reserve Texts

|  | No. Texts | No. Words (average) | Sample |
|---|---|---|---|
| FFR Announcements | 181 | 400 | 2000–2020 |
| FOMC Minutes | 170 | 6809 | 2000–2020 |
| Chairman Speeches | 425 | 2931 | 2000–2020 |
| Total | 776 | 3213 | 2000–2020 |

*Sources*: The Federal Reserve Board of Governors and FederalReserve.gov archives.

**2.2 Methodology**

To quantify instances of Fed communication, we employ text-based measures assessing uncertainty and sentiment, utilizing a range of custom dictionaries. The methodology encompasses sentiment scoring, simple word-counting procedures, and topic modeling techniques applied to the collection of texts.

First, we employ word counting procedures using distinct dictionaries: a finance-specific sentiment dictionary (Loughran and McDonald, 2011; hereafter, LM), an FSS[6] dictionary (Correa et al., 2021), a UMP dictionary, and UMP and COVID-19 lexicons.

Second, we utilize sentiment scoring based on the LM dictionary to proxy for sentiment and contextual uncertainty in the Fed's communications. Multiple sentiment scores and polarity indicators are built using general (NRC, SentiWords, Hu&Liu, Jockers) and specialized (FSS, UMP) dictionaries. Following Loughran and McDonald (2016), we incorporate valence shifters to capture nuances in sentiment, such as negation (e.g., "not good"), amplification (e.g., "very good"), or de-amplification (e.g., "somewhat good"). This approach allows us to more accurately assess the sentiment conveyed in Fed communications, recognizing that context significantly alters the meaning and intensity of sentiment-laden words.

Third, we employ topic modeling for unsupervised machine learning analysis, specifically the Latent Dirichlet Allocation (LDA) algorithm (Blei et al., 2003). This technique allows us to extract and examine thematic content in the Fed's communications, comparing it to economic and financial developments. LDA treats documents as random mixtures over latent topics, where each topic is characterized by a distribution over words. This probabilistic approach enables us to identify the main themes in Fed communications without imposing predefined categories.

---

[5] The full dataset can be accessed at github.com/CSSEGISandData/COVID-19
[6] Correa et al. (2021) capture movements in financial cycle indicators and sentiments conveyed in financial stability reports.

The dictionary capturing the sentiment about UMP measures merges two dictionaries: Erasmus and Hollander (2020), which focuses on forward guidance and quantitative measures, and Henry (2008), which focuses more on the regulatory context, structural attributes, and the dual informational-promotional role of earnings press releases.

The lexicon that captures communications regarding the UMP measures (UMP terms) collects words related to UMP policies from Fed communications using topic modeling and Bag-of-Words techniques (Table 2).

**Table 2.** Unconventional Monetary Policy Lexicon

| | | | |
|---|---|---|---|
| asset purchases | depreciation pressure | market disrupt | risk premium |
| helicopter | direct lending | market functioning | securities purchases |
| QE | ELB | monetary base | stagflation |
| securities purchases | foreign exchange reserve | monetary stimulus | support |
| balance sheet | forward guidance | money supply | support liquidity |
| business support | funding | negative policy | supporting corporat |
| credit facilit | insolvency | negative rate | swap line |
| credit impair | intervention | NIRP | unconventional |
| deferral | lending facilit | quantitative easing | ZLB |
| deflation | lower bound | relaxing regulatory | |

*Notes*: This dictionary primarily includes words and their root forms extracted from Fed communications using topic modeling and Bag-of-Words techniques (see Benchimol et al., 2022).

The lexicon that captures communications regarding COVID-19 compiles relevant keywords related to the pandemic, identifying virus-related content (word frequency) in the Fed's communications (Table 3). By counting the number of COVID-19-related words, we can estimate the attention paid to the pandemic at each point in time.

**Table 3.** COVID-19 Lexicon

| | | | | |
|---|---|---|---|---|
| acute | elderly | infect | pandemic | severe acute |
| cases | emergency | infection | pneumonia | sickness |
| confin | epidem | infection rate | quarantine | spreading |
| contagio | epidemic | lockdown | relief | syndrom |
| corona | hcov | mask | reproduction rate | testing |
| coronavirus | health | medical | respirator | vaccin |
| covid | hospital | morbid | respiratory | virus |
| death | hubei | morbidity rate | sars | wave |
| disabilit | human | mortal | sars cov | wuhan |
| disease | illness | ncov | sarscov | |
| disorder | inception rate | outbreak | sars-cov | |

*Notes*: This dictionary includes words and root words related to the COVID-19 that appeared in both media sources (e.g., Google Trends search queries) and recent Fed communications (Bag-of-Words).

For sentiment analysis, the LM dictionary, designed for assessing sentiment and uncertainty in financial text and widely used in the finance and economics literature (Loughran and McDonald, 2016), is complemented by commonly used sentiment dictionaries in text mining literature (Jockers, NRC, Hu&Liu). Valence shifters are incorporated to capture nuances in sentiment. Sentiment scores are calculated using a formula that accounts for the positive and negative words in the corresponding dictionaries for the text, such that:

$$Score_t = \frac{Positive_t}{Positive_t + Negative_t}, \quad (1)$$

where $Positive_t$ ($Negative_t$) is the total number of words classified as positive (negative) in the corresponding dictionary for text $t$.

Two types of sentiment indicators based on the LM dictionary are constructed: a standard score measure and a polarity measure that includes neutral, positive, negative, very positive, or very negative sentiment, considering the words preceding and following the target word.

## 3. Results
**3.1 Sentiment Analysis**

This section presents sentiment scores, providing insights into the Fed's response to the evolving developments during the COVID-19 pandemic, marked by a notable decline in sentiment scores and an increase in uncertainty-related words in the first quarter of 2020.

Figure 1 shows sentiment scores within FFR announcements, revealing a downturn from January to April 2020, coinciding with the onset of the COVID-19 crisis. The depicted sentiment trends reflect an overall pessimistic view during the early months of the pandemic, but indicate a more optimistic sentiment by the third quarter of 2020 (Panels A1, A3, A6, and A8). This timeframe aligns with heightened contextual uncertainty (Panel A2), denoting the number of words indicative of uncertainty scaled by text length.

Utilizing the LM dictionary, the polarity indicator (Panel A3) reveals a decline but exhibits a more optimistic sentiment for 2020:Q3. FFR announcements serve as a summary of the current state of the economy and monetary policy decisions. Figure 1 shows that the impact of the shock that started in January 2020 persisted until April 2020. The sharp decline in sentiment observed in 2020:Q1-Q2 coincides with the rise in contextual uncertainty that lasted from 2020:Q1-Q3. These trends mirror the global spread of COVID-19 and its anticipated spillover effects on the US economy.

Figure 1 presents a comparative analysis of sentiment indicators, revealing nuanced dynamics in the polarity indices derived from Jockers, NRC, SentiWords, Hu&Liu, and our LM-based polarity index. Notably, the Jockers, NRC, and SentiWords indices appear less informative as they are based on general interest dictionaries, thus ignoring the specifics of economics and finance, while the Hu&Liu polarity index exhibits similar dynamics to our LM-based polarity index.

**Figure 1.** Sentiment Scores

Panel A: FFR Announcements

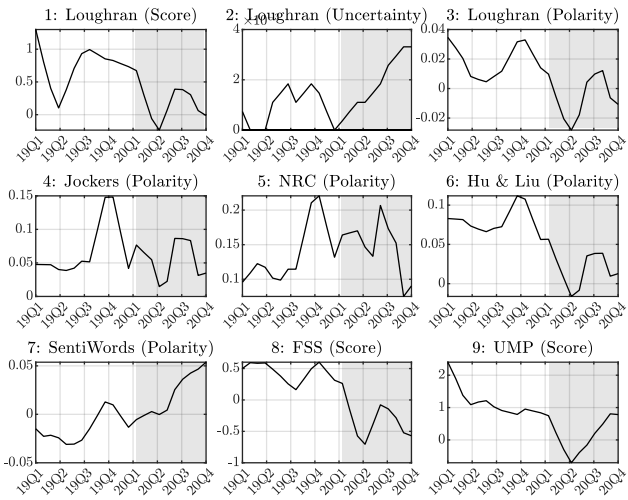

Panel B: FOMC Minutes

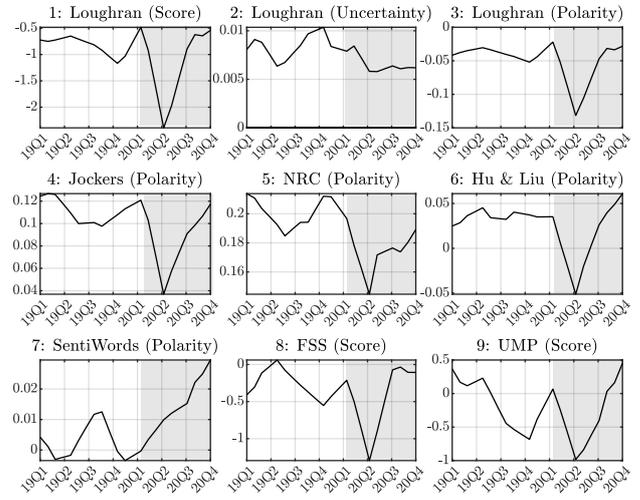

Panel C: Chairman Speeches

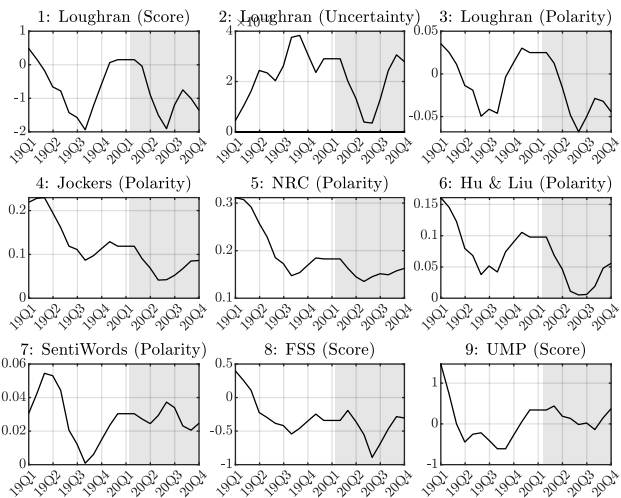

Panel D: Main Fed Communications

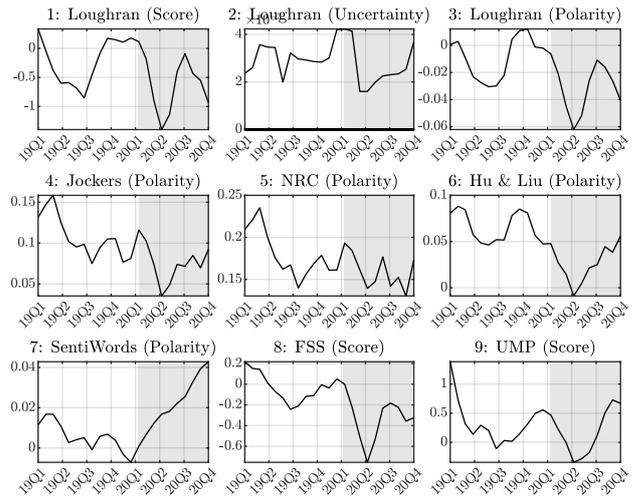

*Notes*: Solid black lines represent sentiment score values. The shaded area represents the COVID-19 crisis.

The broad-coverage polarity index SentiWords captures a noteworthy surge in sentiment within FFR announcements from the onset of the COVID-19 crisis. This suggests a potential shift in the Fed's communication strategy compared to other crises,[7] possibly reflecting a deliberate attempt to convey confidence amid unprecedented economic uncertainty.

The FSS has experienced a decline into negative territory since 2019:Q4, indicating a prevalence of negative financial stability-related words in FFR announcements during this period, which preceded the actual outbreak of COVID-19 in the US. This suggests that concerns about financial stability were emerging even before the pandemic began to significantly impact the US economy.

---
[7] The SentiWords polarity index declines sharply for the GFC and dot-com crises. The full sample results are available upon request. See also Benchimol et al. (2020).

Before the COVID-19 crisis, the sentiment related to UMP was notably positive, reflecting the Fed's plan to gradually reduce the size of its balance sheet. However, with the onset of the COVID-19 crisis, emergency policies were accompanied by communications with a more negative tone, contributing to a surge in financial uncertainty (Section 6).

Figure 2 shows a comparative sentiment analysis across the entire sample covering the dot-com, GFC, and COVID-19 crises. The figure presents sentiment trajectories during the three major economic crises of the past two decades, allowing us to compare how the Fed's communication sentiment evolved during each period of economic stress.

**Figure 2.** A Tale of Three Crises: Sentiment.

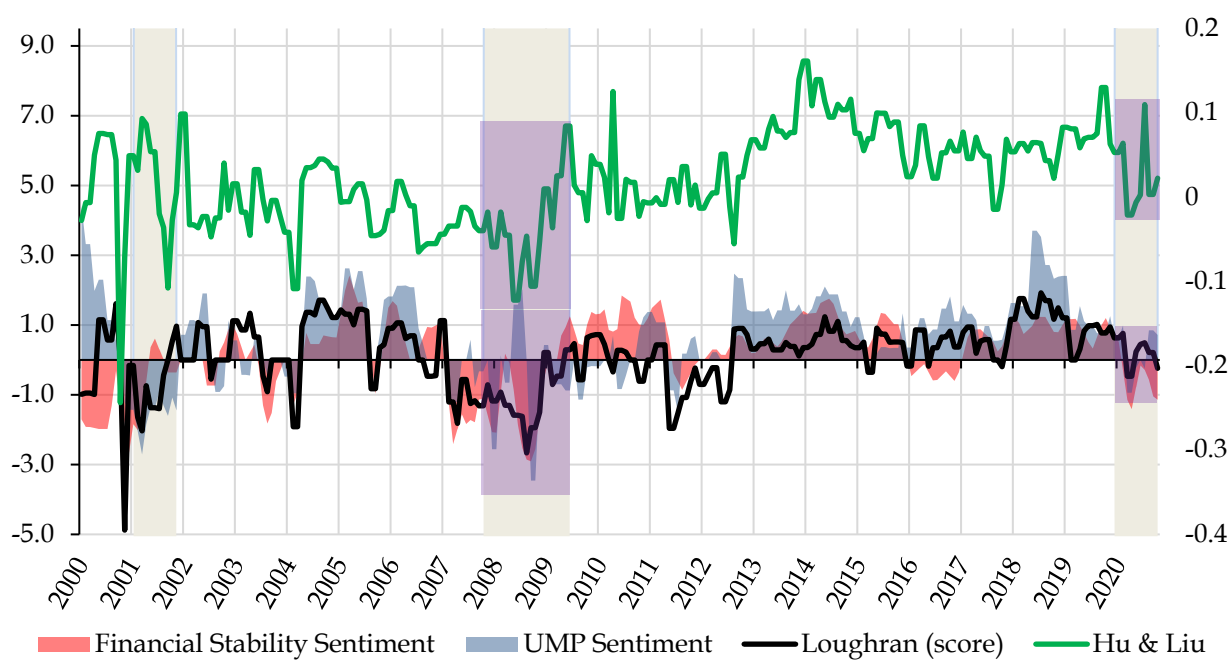

*Notes*: The vertical shaded areas represent the NBER recession periods. Solid lines represent sentiment scores computed from FFR announcements. The different colored lines represent different sentiment measures as indicated in the legend.

The figure presents a more stable sentiment trajectory during the COVID-19 crisis compared to the sharp declines seen in previous crises, suggesting a different communicative approach by the Fed during the recent pandemic. This could reflect the Fed's learning from previous crises and its enhanced ability to manage communication during periods of economic stress. It could also indicate that the Fed was attempting to project greater stability and confidence despite the severity of the economic shock caused by the pandemic.

Figure 2 shows that the FSS experienced a marked deterioration before the GFC and the dot-com crisis, contrasting with its significantly positive status preceding the onset of the COVID-19 pandemic. In the context of sentiment polarity indices, the Hu&Liu index improved from the GFC to the COVID-19 crisis, with a less pronounced decline during the latter than during the former. In contrast, LM's sentiment index did not exhibit improvement between crises, but it experienced a milder decline during the COVID-19 crisis than in previous crises. Furthermore, the volatility of these indicators was less conspicuous during the COVID-19 crisis than during the GFC and dot-com crises. While Figure 2 provides insights into the dynamics of the three crises, it also vividly depicts the distinct trajectories of Fed communication policies across these periods.

Figure 1 (Panel B) illustrates sentiment scores in FOMC minutes, with an increasing sentiment trend during the early phase of the COVID-19 crisis. This positive trend may reflect the Fed's strategic use of reassuring language to mitigate economic concerns during this period. The dynamics of nearly all sentiment indicators show a sharp deterioration in 2020:Q2, a trend more pronounced for the FOMC minutes than for the FFR announcements.

As a possible indication of the disparity between the depiction of the ongoing economic scenario in FFR announcements and the in-depth discussions and provisional solutions to the COVID-19 crisis found in the minutes, the LM score and polarity indices underwent a pronounced decline in sentiment related to financial uncertainty from January to April 2020.

The sentiment score related to UMP declined until 2020:Q2, followed by a sharp upturn leading to a positive trajectory. This shift aligns with the more positive language observed in FOMC minutes concerning UMP measures undertaken during the COVID-19 crisis. Figure 1 (Panel B8) also highlights that policymakers perceived a significant financial stability risk in 2020:Q2, according to FOMC minutes, albeit less severe than during the GFC.[8]

The SentiWords index, known for its extensive coverage, reveals an intriguing pattern of increasing sentiment from the onset of the COVID-19 crisis. This trend may be attributed to the Fed's communication strategy, employing more positive language to reassure and pacify economic agents. In contrast, the SentiWords indicator sharply declined to historically low levels during the GFC, suggesting a different approach to crisis communication during the COVID-19 pandemic.[9]

Figure 1 (Panel C) provides sentiment indicators for the Fed Chairman's official speeches, showing greater volatility but a milder decline in sentiment than in the FOMC minutes and FFR announcements. This suggests that the Chairman's speeches may have focused more on managing expectations during the pandemic. The flexibility of speeches compared to the straightforward and more supervised content of the minutes and announcements explains this result.

---

[8] The full sample results are available upon request. See also Benchimol et al. (2020).
[9] The volatility in sentiments conveyed by Fed Chairman's speeches compared to FFR announcements and FOMC minutes is discussed in Benchimol et al. (2020).

Despite a worsening economic situation and declining sentiment from February 2020 onward, there is a concurrent decrease in contextual uncertainty. This apparent contradiction might reflect the Chairman's efforts to reduce perceived uncertainty through clear communication about the Fed's policy responses, even as the economic outlook deteriorated.

However, the relatively small sample size of Chairman speeches during the COVID-19 crisis may affect the volatility of sentiment indicators presented in Figure 1 (Panel C), making them less erratic than those for the GFC.

Similar to FFR announcements and FOMC minutes, the expansive scope of the SentiWords dictionary captures the distinct communication policy employed by the Fed during the COVID-19 crisis, evident at least until 2020:Q4. The LM dictionary also reveals a decline in the use of uncertainty-related words in Chairman speeches following the COVID-19 outbreak in China, potentially reflecting a deliberate communication strategy to project confidence and reduce market uncertainty.

Figure 1 (Panel D) presents a consolidated view of the Fed's communication strategies during critical periods by aggregating FFR announcements, FOMC minutes, and Fed chair speeches into a single corpus for sentiment analysis. This comprehensive approach allows us to identify overall trends in the Fed's communication sentiment across different channels.

In Figure 1 (Panels C and D), the sentiment scores show a marked decrease in uncertainty sentiment from January to April 2020. This period correlates with the Fed's implementation of liquidity measures to stabilize the financial market. Figure 1 suggests that the Fed's communication was consistently tailored across various types, such as FFR announcements, FOMC minutes, and Chairman speeches, to address the crisis effectively.

By consolidating all communication types within our dataset, this comprehensive overview of the Fed's communications reveals that the substantial decline in uncertainty sentiment was primarily driven by minutes and speeches. The FSS also exhibited a notable decline from January to April 2020, correlating with liquidity measures taken by the Fed to stabilize market volatility and uncertainty, as elaborated in Section 6.

The findings presented in Figures 1 and 2 consistently lead to a compelling conclusion: During the COVID-19 crisis, the Fed conveyed specific sentiment patterns across all communication types—FFR announcements, FOMC minutes, and Chairman speeches. Moreover, these results underscore the distinct sentiments prevailing during the COVID-19 crisis compared to the GFC, suggesting an evolution in the Fed's crisis communication approach over time.

This section delves into the noteworthy sentiment deterioration experienced from January to April 2020. Subsequently, sentiment rebounded into positive territory, with certain measures surpassing pre-crisis levels. By examining the pronounced temporal variations in sentiment, we observe significant changes in the Fed's communication patterns during the early stages of the pandemic, followed by a shift toward more positive sentiment as the Fed implemented its policy responses.

Statistical evidence confirms milder sentiment deterioration during COVID-19. The average magnitude of sentiment breaks was 0.157 for COVID-19 compared to 0.213 for the GFC, with statistically significant differences.[10] In addition, break points in sentiment occurred approximately a month earlier in the COVID-19 crisis than in previous crises, suggesting more rapid Fed communication adaptation. The analysis reveals that break magnitudes differ systematically across communication channels, with FSS showing larger breaks in FFR announcements (0.217) than in speeches (0.156). Post-break sentiment trajectories exhibit distinct patterns across crises, with COVID-19 showing more rapid mean reversion, supporting the hypothesis of enhanced crisis management.

### 3.2 Topic Modeling

This section employs topic modeling to elucidate the underlying themes influencing Fed communication. The analysis, facilitated by the Latent Dirichlet Allocation (LDA) algorithm—a widely recognized unsupervised machine learning technique—identifies and quantifies prevalent themes in the text corpus, offering insights into the Fed's real-time assessments of economic and financial risks.

The examination reveals that, during the initial stages of the COVID-19 pandemic, themes related to policy intervention assumed prominence, overshadowing discussions on inflation expectations and financial stability. This focus on policy intervention in the Fed's communication during the COVID-19 pandemic serves as a distinctive feature compared to the communication strategies employed during the GFC.

Emphasizing that the topics were extracted directly from the text sample, it is crucial to note that the employed topic modeling methodology does not rely on predetermined dictionaries. In contrast to sentiment analysis, this approach represents a less supervised method for discerning word-topic linkages within texts, allowing for the emergence of natural thematic patterns in communications.

The topics extracted from FFR announcements over two decades are visualized in Figure 3, highlighting that policy interventions became a more dominant topic during the COVID-19 crisis, while there was less focus on inflation expectations. Compared to other crises, the discourse on policy interventions garnered increased prominence during the COVID-19 crisis.

Figure 3 illustrates a decline in the significance of the topic of inflation expectations during the COVID-19 crisis, juxtaposed with a heightened importance of the topic of economic growth, encompassing considerations and concerns related to economic growth.

Moreover, Figure 3 offers insights into the relative impact of each crisis on the Fed's communications through FFR announcements. While the dot-com crisis minimally influenced the topics conveyed to the public in FFR announcements, both the GFC and the COVID-19 crises significantly shaped the topics conveyed. This suggests that the severity and nature of these latter crises required more substantial shifts in communication focus. During the European debt crisis, financial stability became a much more prominent topic in the Fed's communications than during the dot-com crisis.

---

[10] T-statistic: 2.76; p value : 0.009.

**Figure 3.** A Tale of Three Crises: Topics

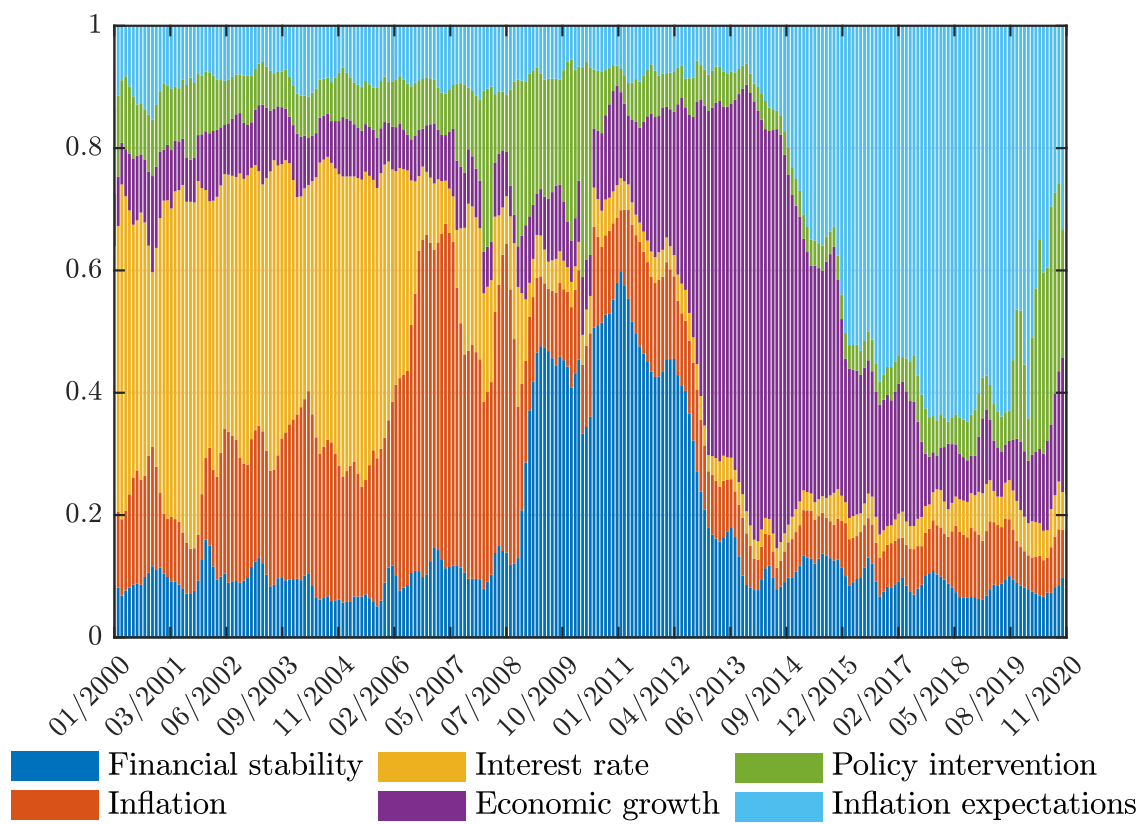

*Notes*: Bars represent the topic probability computed from FFR announcements. For clarity and robustness, we restrict attention to the six most frequently discussed topics. Benchimol et al. (2020) present the topic analysis and words constituting each topic used to generate this figure. The pink bars represent "Policy Intervention," yellow bars represent "Interest Rate," purple bars represent "Economic Growth," green bars represent "Inflation," blue bars represent "Financial Markets," and orange bars represent "Inflation Expectations."

Figure 4 (Panel A) shows the results of topic analysis using FFR announcements, indicating a significant increase in discussions about policy interventions following the COVID-19 outbreak. This suggests a proactive communicative stance by the Fed in response to the emerging crisis, with a clear emphasis on communicating its policy actions to market participants and the public.

The onset of the COVID-19 outbreak coincided with a notable decline in the topic probability related to inflation expectations in FFR announcements. This aligns with the prevailing monetary policy considerations during that period when the focus shifted prominently to policy intervention and addressing the immediate economic fallout from the pandemic.

Additionally, there was a smaller decline in discussions concerning inflation, coupled with an increase in discussions about economic growth within FFR announcements during the COVID-19 crisis. This shift in focus reflects the Fed's priorities during the pandemic, with greater emphasis on supporting economic activity and less concern about immediate inflationary pressures.

**Figure 4.** Topic Analysis

Panel A: FFR Announcements

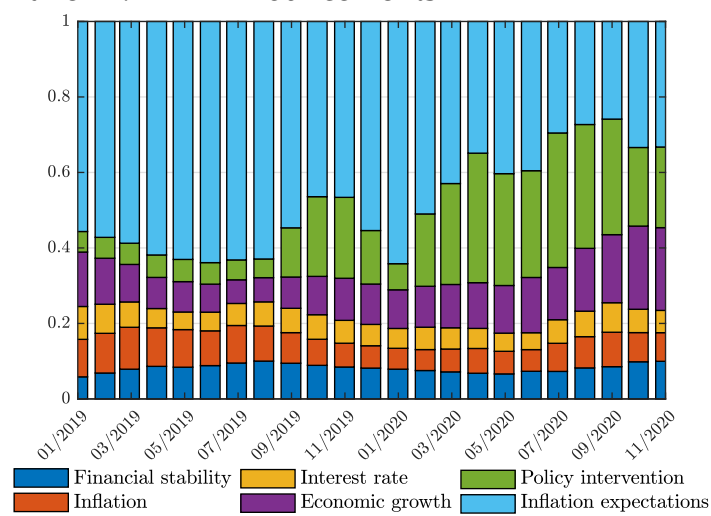

Panel B: FOMC Minutes

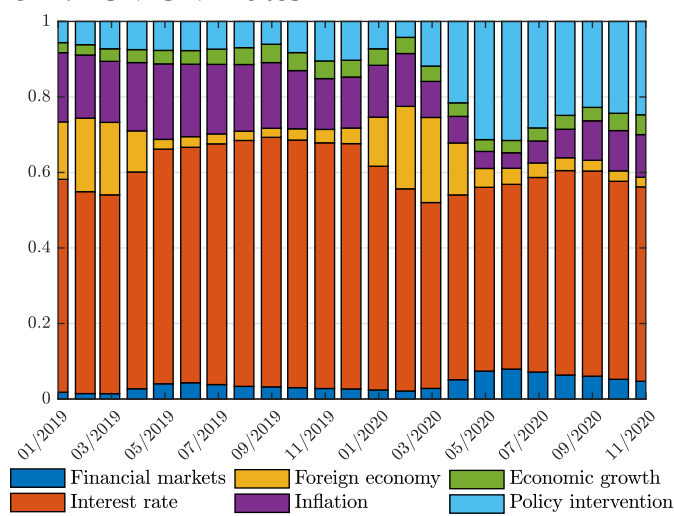

Panel C: Chairman Speeches

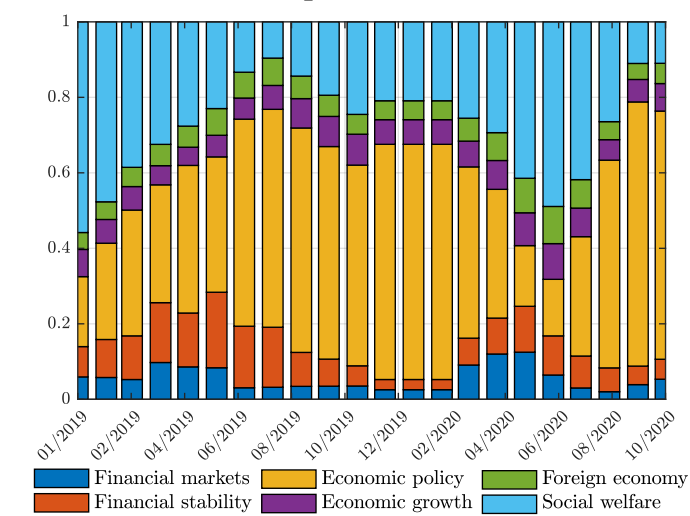

Panel D: Main Fed Communications

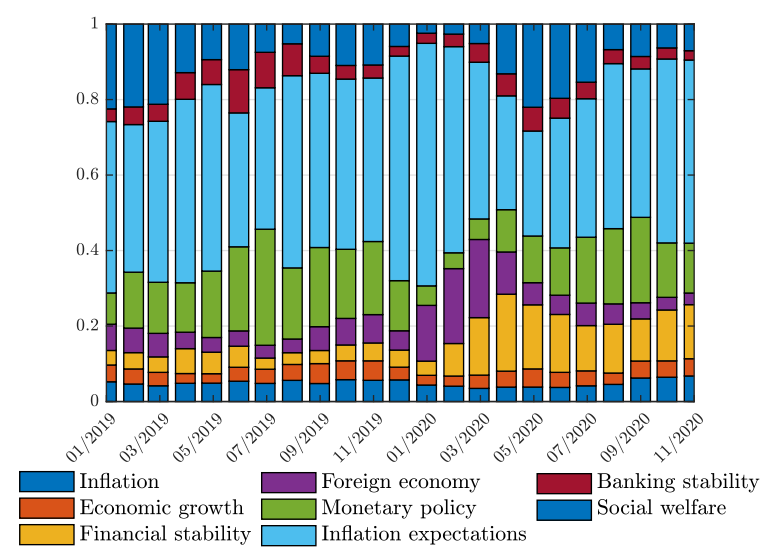

*Notes*: For clarity and robustness, we restrict attention to the six most frequently discussed topics, except for Panel D, considering a broader set of texts. Words constituting each topic are available upon request.

The observed escalation in the topic probability of policy intervention in FFR announcements led to a corresponding reduction in the topic probability related to inflation expectations and, to a lesser extent, to inflation. Despite one of the Fed's primary objectives being the stabilization of prices, this finding suggests that FFR announcements became less tethered to inflation concerns following the COVID-19 outbreak. Furthermore, the slight increase in the topic probability associated with economic growth after the COVID-19 outbreak in China indicates the Fed's apprehension that the pandemic threatened economic growth.

Figure 4 (Panel B) presents results for the topic analysis of FOMC minutes, demonstrating an increased focus on policy interventions and financial markets, indicating the committee's heightened attention to these areas during the COVID-19 pandemic.

Notably, akin to FFR announcements, the FOMC minutes also exhibit an influence of the policy intervention topic, despite the predominant focus on interest rates. The probability of discussing inflation in the FOMC minutes declined following the COVID-19 outbreak, while there was an upsurge in discussions related to policy intervention and financial markets.

The thematic content conveyed by the FOMC minutes underscores a significant augmentation in the coverage of policy intervention and the foreign economy. Intriguingly, the expansion of coverage concerning the foreign economy had commenced even before the onset of the COVID-19 outbreak, while coverage of topics such as inflation and interest rates had already begun to decline. This suggests the Fed was monitoring international developments and potential spillovers before the pandemic fully emerged in the US.

Figure 4 (Panel C) provides a topic analysis of speeches by the Chairman of the Fed, noting a shift in focus toward social welfare concerns after the COVID-19 outbreak, which aligns with broader societal issues in the US, such as education and inequality. This observation highlights that Chairman speeches often address matters unrelated to the Fed's primary objective of price stabilization, delving into areas like education, healthcare, and development economics, encompassing family and labor markets.[11]

The divergence in supervision and audience focus between speeches, intended for broader audiences, and traditional FFR announcements and FOMC minutes, which concentrate more on inflation and output growth, may explain the heightened discussions surrounding social welfare issues. A noticeable uptick in discussions related to economic policy is also observed following June 2020. While this could be attributed to COVID-19 spillovers, the influence of the approaching US election cannot be dismissed. Economic policy considerations had already commanded attention in most Chairman speeches before the COVID-19 outbreak.

Figure 4 (Panel D) offers a comprehensive analysis of the Fed's communication strategies by combining FFR announcements, FOMC minutes, and Fed chair speeches into a single corpus for topic modeling. Considering the substantial number of texts analyzed and their distinct characteristics,[12] more topics are considered for the

---

[11] The most frequently used words and word fragments (root words) in the context of the topic of social welfare are *communiti, economi, educ, work, develop, research, busi, job, peopl, help, opportun, import,* and *family*.

[12] The Fed does not communicate similarly or on similar topics across its various channels, which include FFR announcements, FOMC minutes, and speeches by the Fed chairman. Different topics are emphasized and communicated in varying ways across these channels. This lack of consistency in communication

aggregate topic modeling. These encompass most of the topics outlined in Figures 3 and 4 (Panels A to C).

Notably, the monetary policy topic incorporates references to UMP and unemployment, aligning with the Fed's dual mandate. Despite this, the primary focus on inflation expectations persists, a trend possibly reinforced by the challenges posed by the COVID-19 crisis and concerns regarding long-term interest rates.

Figure 4 (Panel D) highlights three prominent topics that swiftly gained prominence. Initially, there was an escalation in Fed communication concerning the foreign economy as the COVID-19 pandemic extended beyond China to the global stage (01-03/2020).

Subsequently, discussions on financial stability gained momentum amid growing fears about the impact of the COVID-19 crisis on the financial system (03-05/2020). Lastly, there was an upswing in Fed communication related to social welfare, corresponding to the perceived necessity for additional relief plans from the government and the Fed (04-07/2020). While discussions on conventional monetary policy, particularly average inflation targeting, and UMP witnessed a decline at the pandemic's onset, they experienced an upward trajectory thereafter (06-10/2020).

In sum, disparities in the content and timing of Fed communication are evident across the three discussed crises. Unlike the GFC and dot-com crises, the COVID-19 crisis prompted a shift in the focus of communications from discussions on inflation expectations to considerations of policy intervention.

Notably, the topic of policy intervention took greater precedence in communication during the COVID-19 crisis compared to the GFC and dot-com crises. This suggests that policymakers not only implemented distinct policy interventions but also approached the discussion of these interventions differently across the crises. The content, sentiment, and timing of these communications were contingent on the nature of the crisis, with the COVID-19 pandemic requiring a particularly rapid and extensive policy response.

To quantify the differences in topic distributions observed across crises, we employ two complementary metrics: (1) topic coherence scores using Normalized Pointwise Mutual Information (NPMI) to evaluate semantic quality, and (2) Kullback-Leibler (KL) divergence to measure cross-crisis differences in topic distributions.

For topic coherence, we calculate NPMI scores for each topic following the approach of Röder et al. (2015):

$$NPMI(w_i, w_j) = \left[log(P(w_i, w_j)/P(w_i)P(w_j))\right]/\left[-log(P(w_i, w_j))\right] \qquad (2)$$

where $w_i$ and $w_j$ are words in the same topic, $P(w_i)$ and $P(w_j)$ represent the probabilities of individual words $w_i$ and $w_j$ occurring in the corpus, and $P(w_i, w_j)$ represents the

---

approaches has implications when considering aggregating these texts regarding the number of topics to assume.

probability of both words co-occurring in the corpus. Higher NPMI scores indicate greater semantic coherence.

For cross-crisis comparisons, we calculate the KL divergence between topic distributions:

$$KL(P_{COVID} || P_{GFC}) = \sum_i P_{COVID}(i) \times log(P_{COVID}(i)/P_{GFC}(i)) \tag{3}$$

where $P_{COVID}(i)$ and $P_{GFC}(i)$ represent the probability of topic *i* during the COVID-19 and GFC periods, respectively. The Policy Intervention topic exhibits both the highest coherence during COVID-19 (0.380) and the largest divergence from previous crises.[13] This constitutes statistical evidence for a significant shift in communication focus during the pandemic, with more consistent and cohesive messaging about policy actions than in previous crises. Similarly, the Inflation Expectations topic shows significantly reduced prominence during COVID-19 compared to the dot-com crisis,[14] confirming the observation that monetary policy communications became less tethered to inflation concerns during the pandemic.

## 4. Unconventional Monetary Policy

This section investigates the correlation between the Fed's communication, its actions, UMP, and the impact of the COVID-19 pandemic. Figure 5 juxtaposes UMP terms with UMP measures as indicated by Fed balance sheet changes, providing insight into the timing relationship between Fed communication about UMP and actual implementation of such policies.

Figure 5 shows that the Fed communicated more extensively regarding UMP during the GFC than during the COVID-19 crisis. However, the crucial element lies in the timing. Unlike the GFC, where a four-month delay existed between UMP actions (July 2008) and UMP communication (November 2008), the Fed acted and communicated swiftly during the COVID-19 crisis, ensuring better coordination between communication and actions. This improved synchronization likely reflects lessons learned from the GFC about the importance of clear and timely communication when implementing unconventional policies.

Moreover, Figure 5 reveals that the Fed's communications on UMP during the COVID-19 crisis, based on a word count of terms from the UMP lexicon (Table 2), align with effective UMP measures that influenced changes in the Fed's balance sheet with some lags. Unlike the GFC, when actions were implemented before communication, during the COVID-19 crisis, actions followed communication, indicating a shift in the coordination of responses. This suggests a more deliberate strategy of preparing markets for policy changes through forward guidance during the pandemic.

Our UMP dictionary proves versatile in identifying periods beyond crises where substantial UMP measures were undertaken to support the US economy. Each peak in

---
[13] KL = 0.834; p value < 0.001.
[14] KL = 0.673; p value < 0.01.

UMP measure-related communication subsequently impacted the Fed's balance sheet shortly after the communication event. Notably, communications in 2013–2014 signaled the cessation of accommodative policies, aligning with the Fed's balance sheet changes, showcasing the UMP dictionary's ability to capture tapering communication strategies.

**Figure 5.** Unconventional Monetary Policy Terms

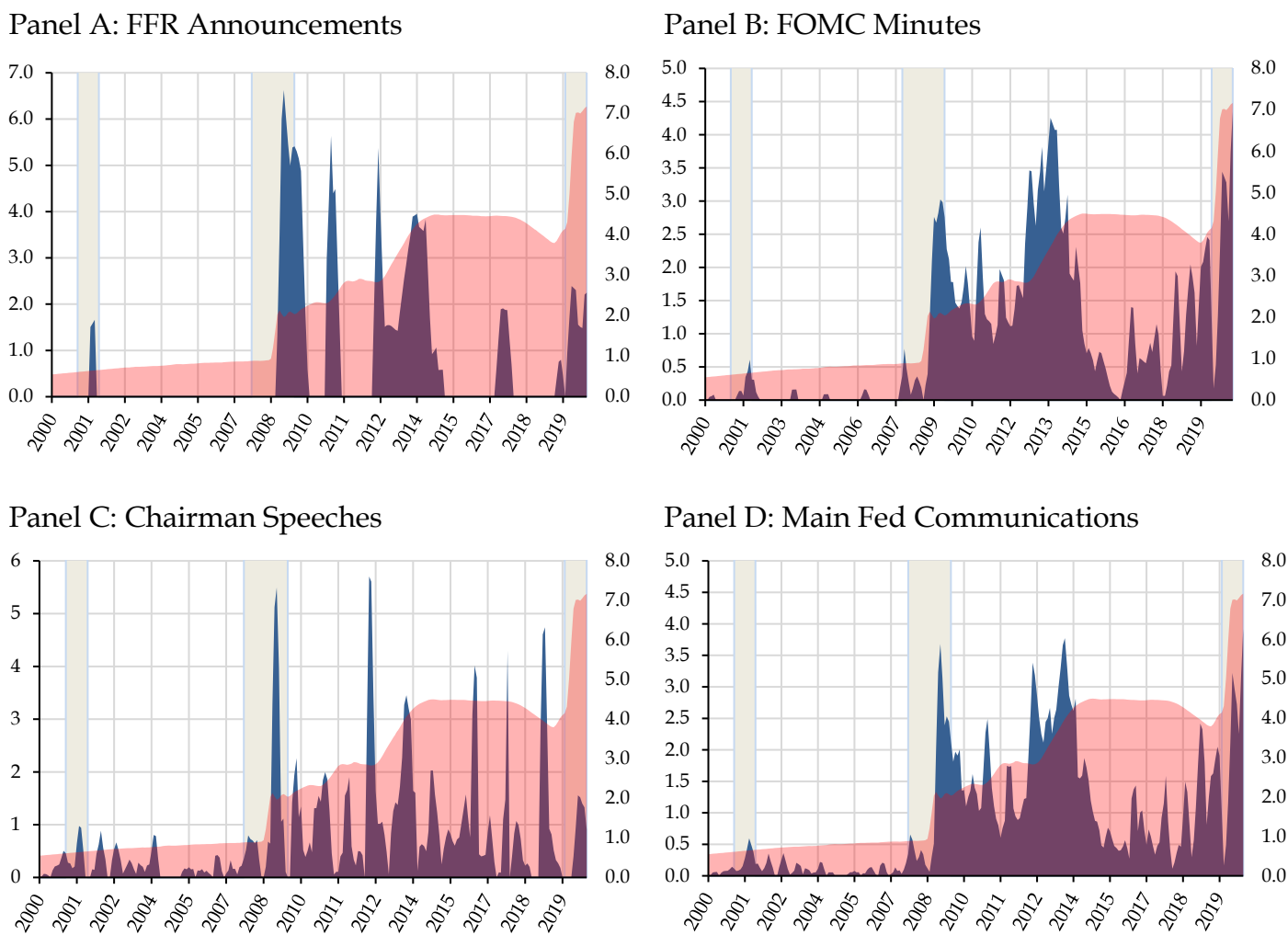

*Notes*: The gray shaded areas represent NBER recession periods. The blue shaded areas represent the word-counting indicator based on the lexicon presented in Table 2 (left axis). The red shaded area represents total assets (minus eliminations from consolidation) in trillions of US dollars in the Fed's balance sheet (right axis).
*Source:* Fed balance sheet data is collected from the Board of Governors of the Federal Reserve System.

In response to economic growth expectations, the Fed initiated a gradual reversal of QE policies, known as "tapering," in 2013. Fed Chairman Ben Bernanke's announcement signaled the reduction of monthly asset purchases if favorable economic conditions persisted, illustrating how communication can be used to prepare markets for policy normalization after periods of unconventional measures.

Panel B of Figure 5 demonstrates that FOMC minutes discussed UMP actions somewhat earlier during the COVID-19 crisis than during the GFC. The frequency of discussions on UMP parallels the tapering discussions conducted by the FOMC in 2013, with an overall higher level observed in late 2020 than during the GFC. This suggests an institutional learning process, with the Fed drawing on its GFC experience to more quickly incorporate discussions of unconventional tools in its deliberations during the COVID-19 crisis.

An intriguing distinction emerges between UMP discussions in FFR announcements and FOMC minutes. Despite the latter being less scripted and more extensive than the former, discussions on UMP in FOMC minutes were more pronounced prior to the COVID-19 crisis than before the GFC or even during the initial phase of the GFC. This observation is corroborated by a comparative analysis presented in panels A and B in Figure 5, suggesting that UMP had become a more routine topic of discussion in FOMC deliberations even before the pandemic struck.

In synthesizing discussions among monetary policy committee members, FOMC minutes consistently incorporate more UMP terms, encompassing discussions and controversies related to potential solutions or policy implementations, than FFR announcements do. The frequent use of such terms dates back to the GFC, indicating a structural shift in the composition and focus of monetary policy discussions. This evolutionary pattern in the Fed's discourse suggests an institutional learning process, wherein unconventional policy tools have become progressively integrated into the standard policy lexicon.

Panel C of Figure 5 provides a word count analysis of UMP-related terms in speeches delivered by the Fed Chairman, revealing distinct communication patterns across different crisis periods.

Figure 5 (Panel C) illustrates that significant communication shocks, such as those observed during the GFC or the third round of monthly purchases of Treasury securities and mortgage-backed securities in September 2012 (third QE), are closely associated with the UMP content found in speeches by the Fed Chairman. Each communication peak coincided with or preceded changes in the Fed's balance sheet dynamics. Peaks that did not impact the Fed's balance sheet were typically linked to forward guidance communications, highlighting the dual role of Chairman speeches in both preparing markets for balance sheet operations and managing expectations about the future path of policy.

Fed Chairman speeches emerge as the primary platform for discussing UMP terms, serving as a critical channel for articulating complex policy innovations to a broader audience. Examining the frequency of UMP terms in speeches delivered before and after each crisis yields intriguing insights. Panel C of Figure 5 reveals an increase in the frequency of UMP terms in speeches following the GFC, dot-com, and COVID-19 crises. Notably, the frequency was more pronounced, and UMP terms occurred earlier during the COVID-19 crisis than in other crises, suggesting a more rapid deployment of the communication toolkit during the pandemic.

Panel D of Figure 5 presents a comprehensive UMP indicator for key Fed communications, aggregating data from FFR announcements, FOMC minutes, and Chairman speeches to provide a holistic view of UMP communication patterns.

The figure highlights distinctions in the timing of UMP communications and actions between the COVID-19 and GFC periods. During the GFC, UMP actions preceded comprehensive communication about these policies, creating a lag in market understanding of the Fed's strategy. In contrast, during the COVID-19 crisis, communications about UMP measures were more synchronized with their implementation, potentially enhancing policy effectiveness through clearer signaling channels.

The emergence of a post-GFC "new normal" is of particular significance, characterized by more frequent UMP communications and actions than during the pre-GFC era. This sustained reliance on UMP tools suggests a fundamental transformation in monetary policy implementation, where previously "unconventional" measures have become standard components of the policy toolkit. This structural evolution reflects the constraints imposed by the proximity to the zero lower bound and the need for alternative policy instruments when conventional interest rate adjustments are limited.

To formalize the lead-lag relationships observed in Figure 5, we implement Granger causality tests within a Vector Autoregression (VAR) framework. For each communication type and crisis period, we estimate:

$$UMP\ Terms_t = \alpha_1 + \sum_{i=1}^{k} \beta_i^1 UMP\ Terms_{t-i} + \sum_{i=1}^{k} \gamma_i^1 Fed\ Assets_{t-i} + \varepsilon_t^1 \qquad (4)$$

$$Fed\ Assets_t = \alpha_2 + \sum_{i=1}^{k} \beta_i^2 Fed\ Assets_{t-i} + \sum_{i=1}^{k} \gamma_i^2 UMP\ Terms_{t-i} + \varepsilon_t^2 \qquad (5)$$

where the optimal lag length *k* is determined using the Bayesian Information Criterion (BIC). We test two null hypotheses:

(1) $H_0: \gamma_1^2 = \gamma_2^2 = \cdots = \gamma_k^2 = 0$, which implies UMP communication does not Granger cause balance sheet changes;

(2) $H_0: \gamma_1^1 = \gamma_2^1 = \cdots = \gamma_k^1 = 0$, which implies balance sheet changes do not Granger cause UMP communication.

We find statistical evidence for a fundamental shift in the Fed's communication strategy between crises. During the GFC, the Fed followed an *action-first* approach, where policy implementation preceded communication, as shown by the significant Granger causality from the Fed Balance Sheet to UMP terms across all communication channels.[15]

In contrast, during the COVID-19 crisis, the Fed adopted a *communication-first* strategy, where communication about potential measures preceded their implementation. This is demonstrated by the significant Granger causality from UMP terms to Fed assets.[16] This shift may suggest institutional learning from the GFC experience and a greater emphasis on forward guidance during the COVID-19 crisis.

---

[15] F test: 8.23 to 12.36, p value < 0.01.

The differentiation across communication channels is particularly noteworthy. Chairman speeches showed the strongest forward-looking property during COVID-19,[17] consistent with their role in signaling upcoming policy changes. FOMC minutes, while still exhibiting the timing shift, showed a less dramatic pattern change, reflecting their function as detailed records rather than active signaling tools.

In summary, communicating about QE and providing forward guidance (UMP) have evolved into the "new normal" for the Fed since the GFC (Bernanke, 2020). This transformation represents a significant institutional adaptation, whereby communication has become increasingly integral to policy implementation, particularly in low interest rate environments. Furthermore, the frequency of UMP terms remained elevated during the COVID-19 crisis compared to preceding crises, indicating the continued importance of these policy tools in the Fed's response to economic disruptions. The statistical significance of UMP-related Granger causality across all communication types quantifies this "new normal."

## 5. COVID-19 Pandemic

This section investigates the utilization of COVID-19 terms presented in Table 3. We analyze these terms alongside UMP terms, contextual uncertainty terms, financial volatility, and new COVID-19 cases to understand how the Fed's communication evolved during the pandemic and how it related to the progression of the health crisis.

Figure 6 delineates the distribution of COVID-19-related terms employed in the Fed's primary communications throughout 2020. The figure reveals a striking pattern: the speeches of the Fed Chairman anticipated the surges in new COVID-19 cases. This observation warrants careful interpretation, given the delayed commencement of testing in the US compared to other countries. Nevertheless, the speeches exhibited a forward-looking perspective, anticipating the potential impacts of the virus on the US economy, even as it originated in China and before it significantly impacted US economic activity.

Notably, the Fed Chairman's speeches serve as a more timely and flexible communication vehicle than FOMC minutes or FFR announcements,[18] allowing them to address emerging issues more rapidly. Chairman speeches cover more on-demand topics and are disseminated more quickly and informally (often broadcast live) than other communication types, allowing for considerations related to public health, politics, or foreign affairs that are less emphasized in other communication formats.

---

[16] F test : 10.54 to 17.82, p value < 0.01.
[17] F test: 17.82, p value < 0.001.
[18] This difference arises from the distinct communication goals and review processes associated with each text: FFR announcements (rigorously reviewed, formal statements), minutes (internal dialogues among committee members), and Chairman speeches (intended for broader audience comprehension, not necessarily focused on interest rate decisions). Our texts ignore questions from the public or journalists.

Figure 6 presents two types of COVID-19 waves: one based on new COVID-19 cases from medical statistics and the other depicting the intensity of COVID-19-related terms in the Fed's communications, based on the COVID-19 lexicon presented in Table 3. Visually, Fed communication about the virus appears to precede new case waves,[19] though we are cautious not to claim any causal relationship between these patterns.

**Figure 6.** COVID-19 and Fed Communication

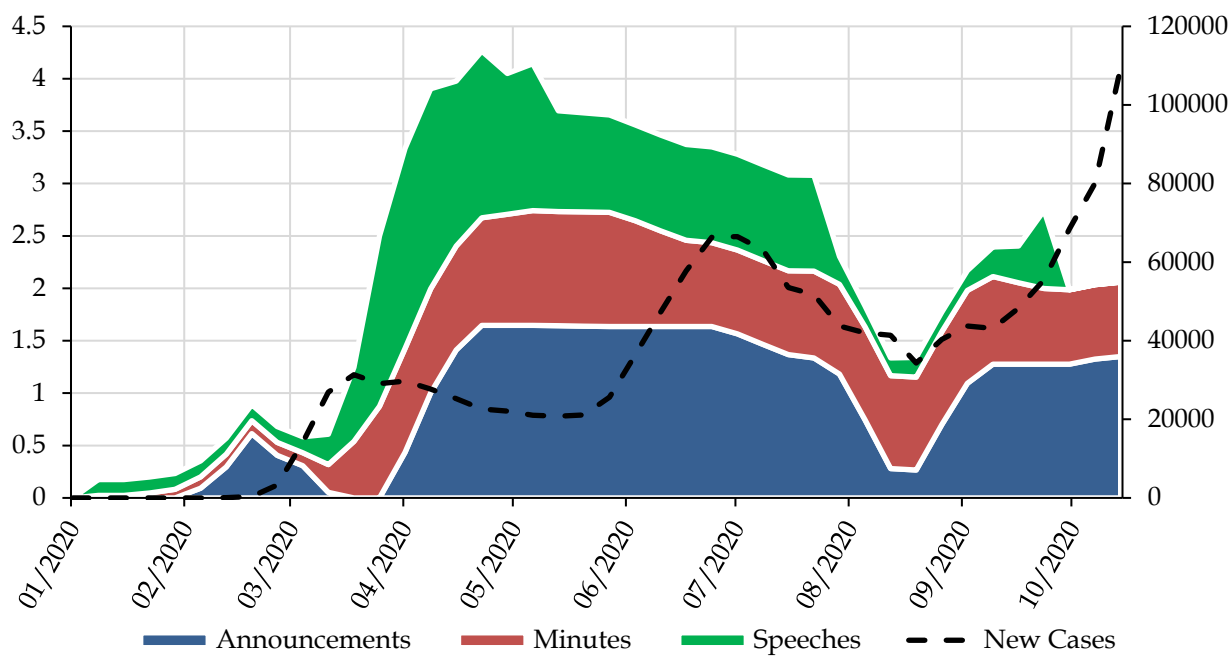

*Notes*: The shaded areas represent the word-counting indicator for each communication type based on the COVID-19 lexicon presented in Table 3 (left axis). The dashed line represents the number of new COVID-19 cases in the US (right axis).
*Source:* Johns Hopkins University Center for Systems Science and Engineering (CSSE).

The COVID-19 virus's magnitude and severity were rapidly acknowledged and communicated to the public by the Fed, primarily through FFR announcements and Chairman speeches. FFR announcements utilized more COVID-19-related terms than other communication types and contributed more to the initial communication wave than speeches, albeit with a slight lag. The decline in the intensity of COVID-19 terms in Chairman speeches in the second quarter of 2020 correlates with a decline in positive sentiment, as reported in Figure 1. This decline aligns with the increased emphasis on social welfare in the Chairman's speeches during this period, as reported in Figure 4. Consequently, both the topics and sentiments of the Chairman's speeches were influenced by the COVID-19 outbreak and its broader economic implications.

Interestingly, the decline in the frequency of COVID-19-related terms after 2020:Q3 explains the SentiWords sentiment increase in Figure 1 (Panels A7, B7, and D7), despite the rise in new cases during that quarter. This pattern suggests that the Fed's communications adopted a more positive tone in the later phases of the pandemic,

---

[19] Granger causality tests also confirm this finding, but given the few observations available, the results are not reported and are available upon request.

potentially to bolster market and public confidence as the economic recovery began to take shape.

In summary, the waves of Fed communications about the COVID-19 crisis anticipated those of new COVID-19 cases. The Chairman's speeches communicated about the first wave of COVID-19 earlier than other communication types (FFR announcements and FOMC minutes), affirming that speeches, being less scripted, allow for quicker communication on emerging issues. This indicates the Fed's early recognition of the severity and magnitude of the COVID-19 pandemic and its related economic implications.

Figure 7 compares new COVID-19 cases in the US with word-counting indicators based on UMP and COVID-19 lexicons, providing insight into the relationship between pandemic developments and the Fed's policy communications.

**Figure 7.** COVID-19 and UMP Terms in Main Fed Communications

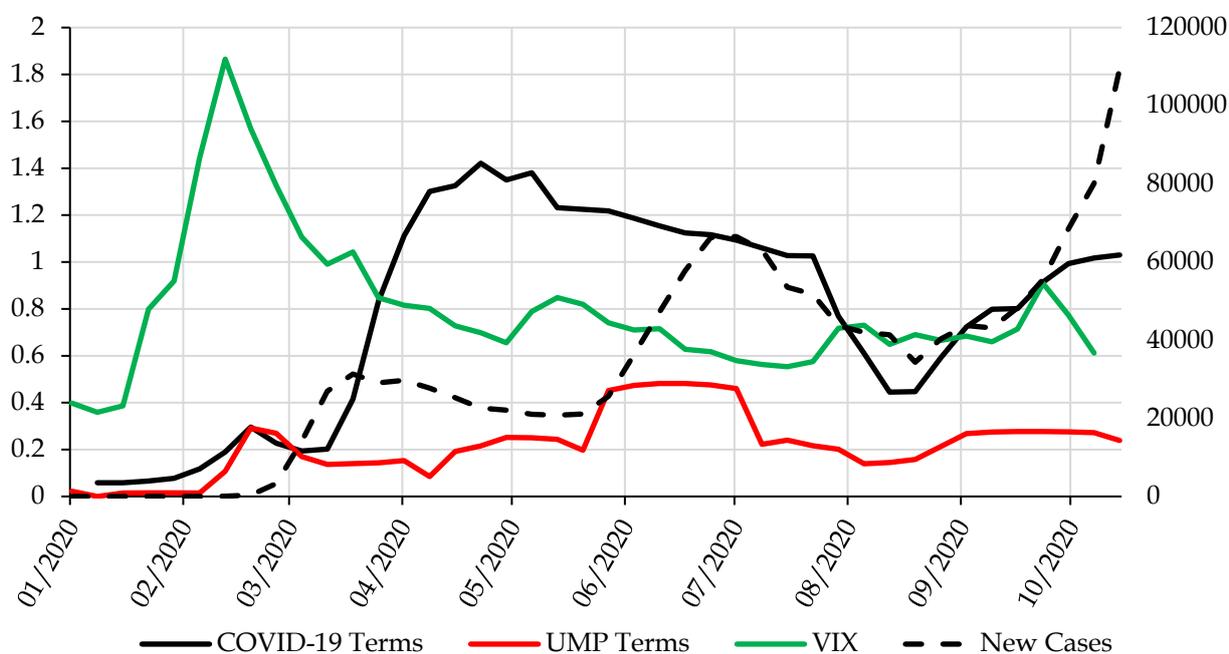

*Notes*: The black line represents the word-counting indicator based on the COVID-19 lexicon presented in Table 3. The dashed line represents the number of new COVID-19 cases in the US (right axis). The green line represents the VIX (volatility index).
*Sources:* Bloomberg and Johns Hopkins University Center for Systems Science and Engineering (CSSE).

The figure also incorporates the VIX to capture financial market volatility during the pandemic, allowing us to examine how the uncertainty created by the virus outbreak related to the Fed's communications about UMP and COVID-19.

In Figure 7, the timing of Fed communications about COVID-19 is compared to the emergence of new COVID-19 cases and financial market volatility, showing that the Fed's use of COVID-19-related terms predated the first wave of cases in the US and coincided with increased VIX values. UMP-related terms and actions in response to the COVID-19 pandemic helped mitigate this volatility, suggesting a stabilizing effect of the Fed's policy interventions on financial markets.

The VIX exhibited a remarkable surge with the onset of the COVID-19 outbreak in China and several other countries, including the US. This spike in market volatility reflected the unprecedented uncertainty created by the pandemic and its potential economic consequences. Notably, mentions of COVID-19 in the Fed's communications preceded considerations of UMP and the subsequent waves of new cases in the US. Between May and July 2020, the Fed extensively communicated about unconventional monetary policies, and during this period, despite a significant increase in new COVID-19 cases, the Fed's communications and actions were associated with a moderation in financial volatility. Following this timeframe, the heightened frequency of UMP-related terms in the Fed's communications as the pandemic worsened might have contributed to stabilizing the VIX, suggesting that clear communication about policy interventions helped reduce market uncertainty.

Figure 8 illustrates our COVID-19 and UMP word-counting indicators alongside the LM dictionary of contextual uncertainty terms, allowing us to examine how pandemic-related communications interacted with expressions of uncertainty in Fed communications.

**Figure 8.** COVID-19, UMP, and Uncertainty in Main Fed Communications

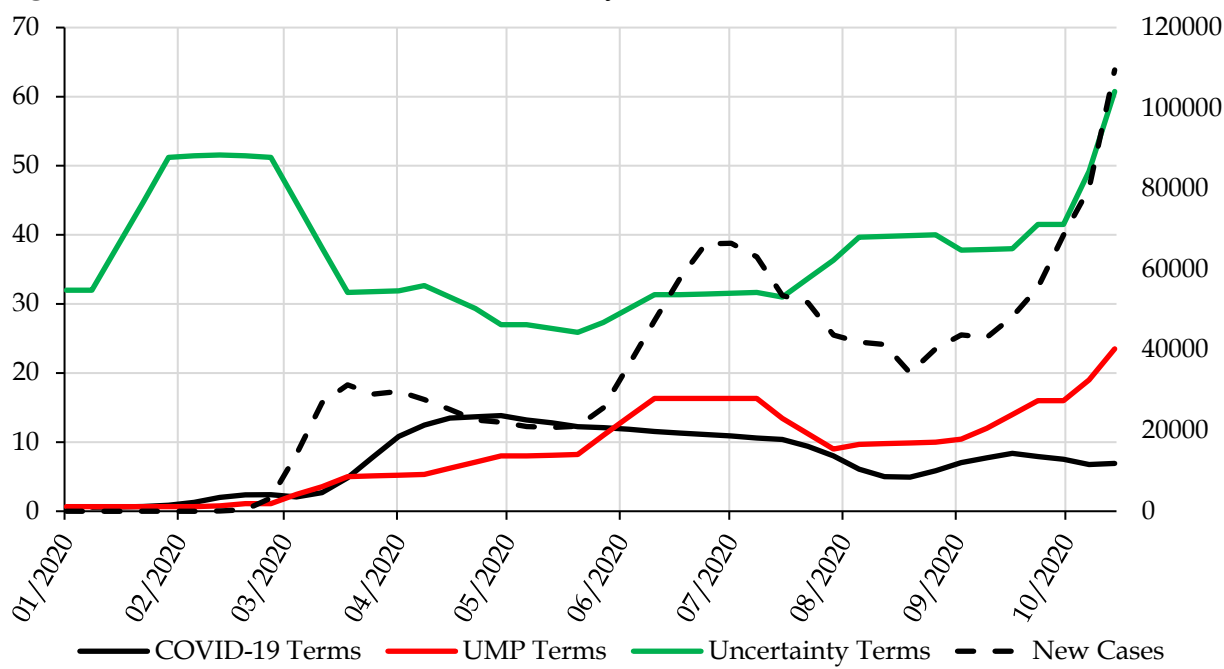

*Notes*: The dashed line represents the number of new COVID-19 cases in the US (right axis). The contextual uncertainty indicator is the number of uncertainty terms, according to LM.
*Source:* Johns Hopkins University Center for Systems Science and Engineering (CSSE).

The uncertainty in Fed communication exhibits a notable correlation with the number of UMP terms found in those communications, representing what we interpret as the "uncertainty effect." New COVID-19 cases, UMP, and uncertainty show significant correlation in the Fed's communications, particularly during the second half of 2020. This correlation was not as prominent at the beginning of the COVID-19 sample period, primarily because the sudden virus outbreak caught everyone by surprise, increasing the frequency of uncertainty-related terms before others. The "uncertainty effect" manifests during crisis periods, necessitating UMP to mitigate market and economic uncertainty.

Figure 8 presents the proactive influence of uncertainty- and UMP-related terms in the Fed's communication concerning COVID-19. The escalation in the utilization of uncertainty terms seems to precede rises in new virus cases. The correlation between contextual uncertainty, defined by the LM dictionary, and UMP-related terms from the UMP lexicon presented in Table 2 is notably positive at 0.44 for the weekly average communications spanning from 2000 to 2020 (i.e., 1090 observations), indicating a consistent relationship between expressions of uncertainty and discussions of unconventional policy measures.

Figure 9 presents our COVID-19 and UMP word-counting indicators alongside the FSS and the number of new COVID-19 cases in the US, allowing us to examine how pandemic-related communications interacted with financial stability sentiment.

**Figure 9.** COVID-19, UMP, and Financial Stability in Main Fed Communications

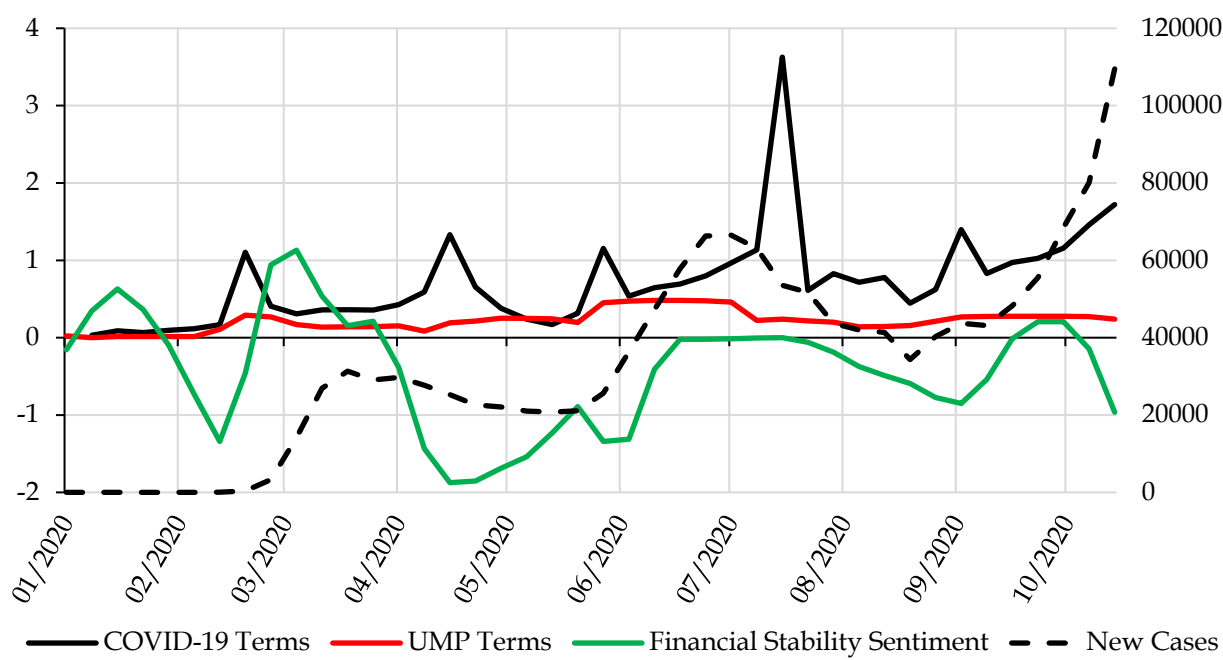

*Notes*: The black line represents the word-counting indicator based on the COVID-19 lexicon presented in Table 3. The dashed line represents the number of new COVID-19 cases in the US (right axis). The financial stability index is rescaled to match scale constraints.
*Sources:* Johns Hopkins University Center for Systems Science and Engineering (CSSE).

Except for the initial phase of the COVID-19 crisis, increases in sentiment associated with FSS align with rises in new virus cases. The diminishing contribution of the Chairman's speeches partially contributes to the decline observed in the FSS toward the end of the sample period.

Figure 9 shows that the decrease in FSS trails a few weeks behind the upsurges in both COVID-19- and UMP-related terms in the Fed's communications. This temporal sequence aligns with theoretical expectations, as discussions and decisions regarding financial stability typically unfold after shocks to financial stability become apparent. The successive deteriorations in FSS preceding the increments in the number of new COVID-19 cases may substantiate the anticipatory nature of the Fed's discussions on stabilization policies.

Following the GFC, Fed communication presaged effective UMP implementations, with the timing and scale of those implementations exhibiting substantial disparities across crises. Figure 9 shows that Fed communication during the COVID-19 crisis similarly foreshadowed waves of new COVID-19 cases. The UMP terms were strategically used to mitigate market volatility and the economic ramifications of COVID-19.

To formally quantify the dynamic relationships between Fed communication indicators and related economic variables during the COVID-19 crisis, we implement a Vector Autoregression (VAR) framework. This approach allows us to establish temporal precedence and measure the magnitude of relationships between communication variables.

We estimate the following five-variable VAR model:

$$Y_t = A_0 + A_1 Y_{t-1} + \cdots + A_p Y_{t-p} + \varepsilon_t \qquad (6)$$

where $Y_t = [COVID\ Terms_t, UMP\ Terms_t, FSS_t, COVID\ Cases_t, VIX_t]$, representing COVID-19 terms frequency, UMP terms frequency, financial stability sentiment, COVID-19 case changes, and market volatility, respectively.

Prior to estimation, we conduct stationarity tests using both Augmented Dickey-Fuller and KPSS tests. COVID-19 cases and VIX are first-differenced to ensure stationarity. We select the optimal lag order using the Bayesian Information Criterion to balance model fit with parsimony.

**Table 4.** VAR Estimates

| Dependent variable | Independent variable | Coefficient | Std error | t-test | p value |
|---|---|---|---|---|---|
| COVID-19 Cases | Constant | 0.281 | 0.027 | 10.2 | 0.00015 |
| COVID-19 Cases | COVID-19 Cases (t-1) | 1.141 | 0.041 | 27.5 | 1.19E-06 |
| FSS | VIX (t-1) | -0.833 | 0.152 | -5.4 | 0.00278 |
| FSS | COVID-19 Cases (t-1) | -0.549 | 0.152 | -3.6 | 0.0154 |
| FSS | FSS (t-1) | 0.669 | 0.21 | 3.1 | 0.0246 |
| FSS | COVID-19 Terms (t-1) | 0.58 | 0.206 | 2.8 | 0.037 |

Table 4 presents significant coefficient estimates from our VAR model, showing how each communication variable predicts subsequent changes in others. We find statistically significant predictive relationships from COVID-19 terms to UMP terms, indicating that discussions of the pandemic systematically preceded discussions of unconventional monetary policy. Similarly, UMP terms positively predict subsequent financial stability sentiment, suggesting that communication about unconventional policy measures contributed to improved financial stability perceptions.

Our analysis further reveals that the contextual uncertainty articulated in the Fed's communications adeptly foreshadowed the waves of COVID-19. Ultimately, the dip in sentiment associated with the Fed's communications on financial stability generally anticipated the surge in the number of new COVID-19 cases. These patterns suggest that the Fed's communications incorporated forward-looking assessments of both the pandemic's progress and its economic implications, reflecting the institution's capacity to integrate diverse sources of information into its policy deliberations and communications.

## 6. Financial Stability

UMP and uncertainty-related terms play pivotal roles in influencing financial stability and volatility. This section delves into examining FSS and contextual uncertainty, explaining their respective dynamics concerning the conventional monetary policy instrument (FFR) and financial market volatility as represented by the VIX.

Figure 10 compares FSS, UMP-related terms, and uncertainty-related terms extracted from FFR announcements with financial market volatility, providing insight into how the Fed's communications about these issues relate to market conditions.

**Figure 10.** Financial Stability and FFR Announcements

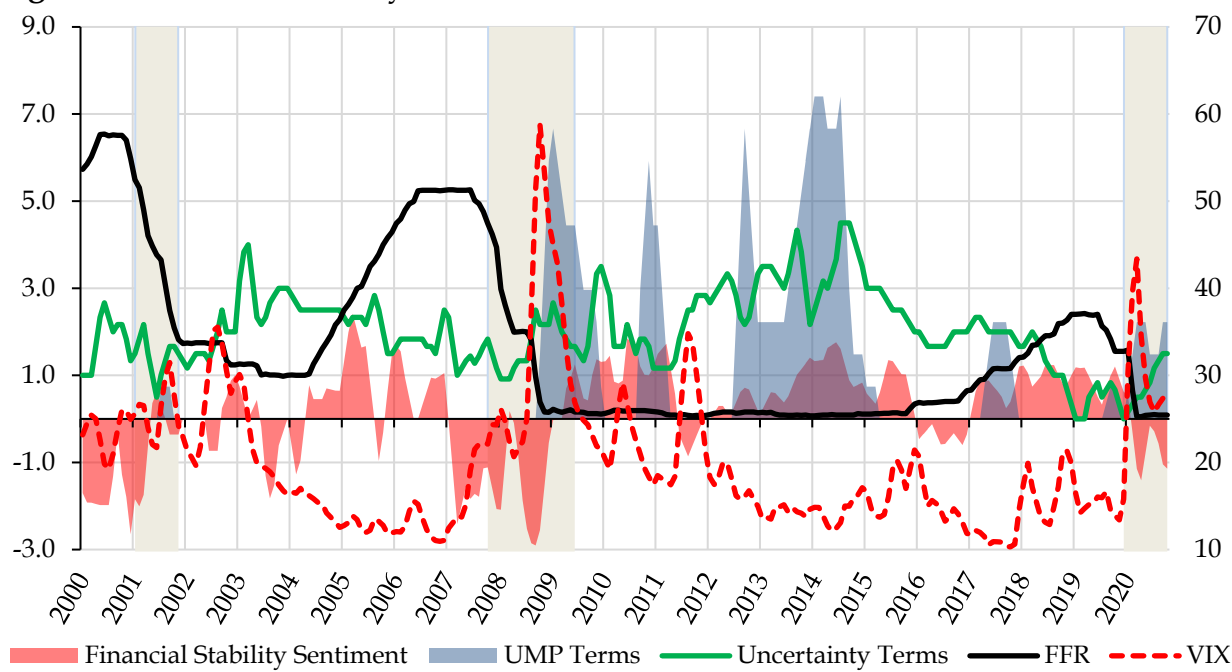

*Notes*: The gray shaded areas represent NBER recession periods. The right axis indicates the VIX level.

Figure 10 shows a reduction in the use of uncertainty-related terms during the pandemic, particularly in comparison to previous crises. This pattern may indicate a deliberate communicative strategy by the Fed to manage market perceptions and reduce perceived uncertainty, even as the economic situation deteriorated.

The Fed appears to have strategically used fewer uncertainty-related terms in its FFR announcements during the COVID-19 crisis than in previous years or crises. This reduction in contextual uncertainty may suggest a communication strategy that aims to mitigate market uncertainty and volatility, especially during periods of heightened concern due to the pandemic. Unlike the dot-com and GFC scenarios, there was no pronounced decline in the FSS during the COVID-19 crisis. This is attributable to the unprecedented and rapidly evolving nature of the crisis and related early interventions, leading to better-managed uncertainties through more coordinated policy responses.

During the dot-com and GFC crises, as well as the COVID-19 crisis, terms associated with UMP played a pivotal role in curbing market volatility. In these critical periods, FFR announcements detailing the implementation of UMP measures had a tangible impact on reducing the VIX. Notably, FSS demonstrated a close association with the FFR level, with a decline in FSS corresponding to a decrease in the FFR, suggesting a consistent relationship between financial stability sentiment and conventional policy adjustments.

Figure 11 mirrors the trends observed in Figure 10, with the focus shifted to FOMC minutes instead of FFR announcements, allowing us to examine how the more detailed policy discussions in minutes relate to financial stability and market conditions.

**Figure 11.** Financial Stability and FOMC Minutes

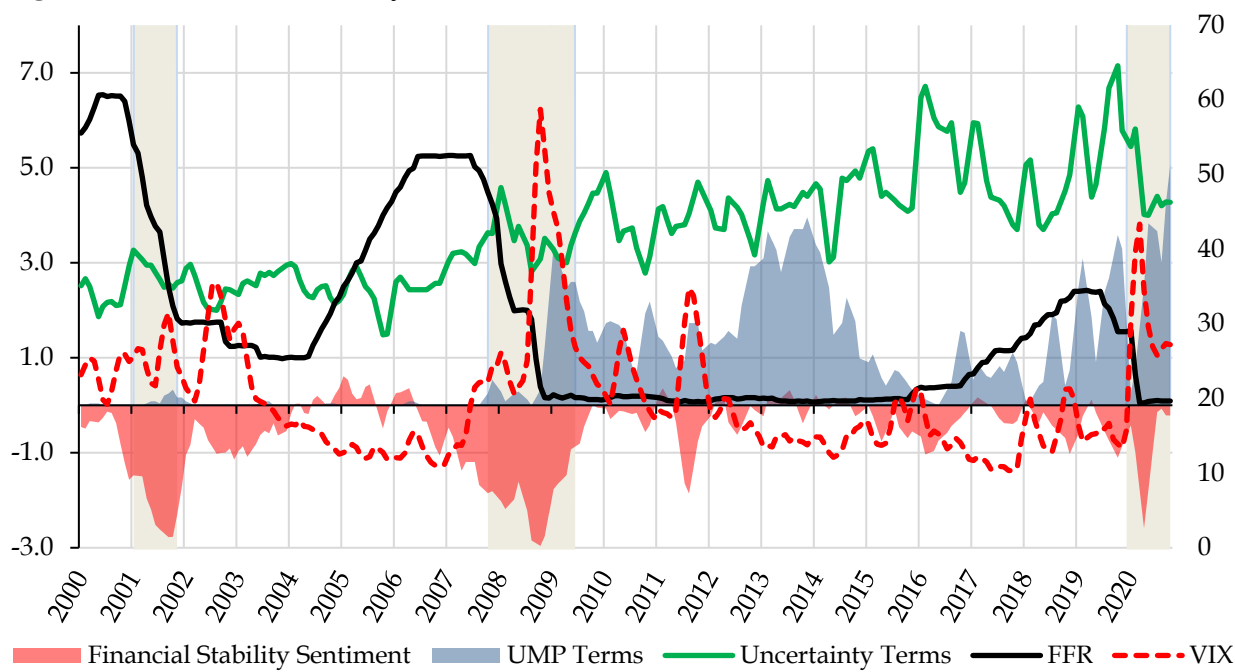

*Notes*: The gray shaded areas represent NBER recession periods. The right axis indicates the VIX level.

There is a discernible uptick in the utilization of uncertainty-related terms in FOMC minutes preceding each crisis. However, an intriguing pattern emerges in Figure 11, akin to the observations in Figure 10, wherein there is a clear inclination to diminish the use of uncertainty-related words during crisis periods. The VIX exhibits a negative correlation with FSS, and both precede UMP terms. This chronological sequence typically signals the implementation of UMP measures to stabilize markets, allaying concerns about financial stability and subsequently reducing financial volatility.

Given that the FOMC minutes offer intricate insights into the monetary policy committee's perspectives on the immediate policy stance and the economic outlook for the US, they serve as an early conduit for conveying FSS and UMP terms compared to FFR announcements. Notably, FFR increases align with elevated FSS levels, except for the period from 2012 to 2015, when UMP communications and actions propelled the FSS upwards, suggesting that unconventional policies can have distinct effects on financial stability sentiment compared to conventional rate adjustments.

Figure 12 juxtaposes Fed Chairman speeches, FFR announcements, and financial volatility, providing insight into how the Chairman's communications relate to market conditions and conventional policy measures.

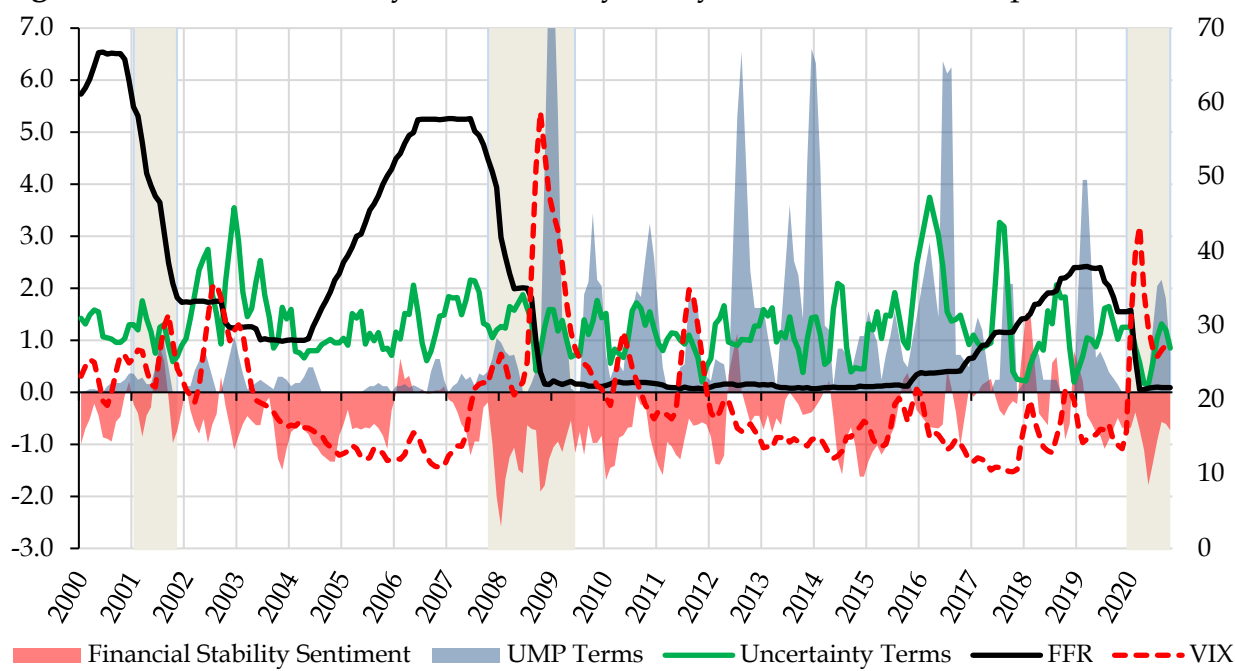

**Figure 12.** Financial Stability and Monetary Policy in Fed Chairman's Speeches

*Notes*: The gray shaded area represents NBER recession periods. The right axis indicates the VIX level.

The FSS expressed in speeches exhibits a lesser predictive capacity for future FFR movements than announcements and minutes do. This may reflect the broader scope and more varied objectives of speeches, which often address issues beyond immediate policy concerns.

Chairman speeches may convey a negative FSS even when the FFR increases, a phenomenon not typically observed in FFR announcements or FOMC minutes (refer to Figures 13 and 14). This divergence could reflect the Chairman's role in providing context and nuance to policy decisions, including acknowledging risks and challenges even as conventional policy tightens.

Drawing a comparison between two pivotal periods—the dot-com to the GFC period (P1) and the GFC to the COVID-19 crisis period (P2)—provides valuable insights. During P1, a scarcity of UMP-related terms in Fed Chairman speeches is observed. In contrast, P2 marks the establishment of a "new normal," marked by a noteworthy correlation between UMP- and uncertainty-related terms, a connection absent in P1.

This correlation extends to the relationship between the VIX and uncertainty-related terms, exhibiting a lesser degree of correlation in P1 than in P2, suggesting a structural change in how uncertainty and unconventional policy are communicated and perceived after the GFC.

Examining Figure 12 reveals a reduction in uncertainty-related terms in Fed Chairman speeches during the COVID-19 crisis in comparison to previous years and crises. However, this contrast is less pronounced in announcements and minutes. The inherent variability in speeches can be attributed to their broad scope and varied objectives, coupled with their somewhat less controlled nature than announcements and minutes.

The communication pertaining to UMP transpired after volatility peaks in both the GFC and COVID-19 crises. Figures 14 and 15 underscore the interventionist approach adopted by the Fed, aligning with the central bank practices in many developed countries. Following each FFR decrease, UMP communication, typically succeeded by actions, serves as a compensatory measure for the central bank's constraint in utilizing the nominal interest rate—the primary policy instrument—at the ZLB.

Figure 13 amalgamates the three communication types employed by the Fed, providing a comprehensive overview of how different communication channels collectively relate to financial stability and market conditions.

**Figure 13.** Financial Stability and Monetary Policy

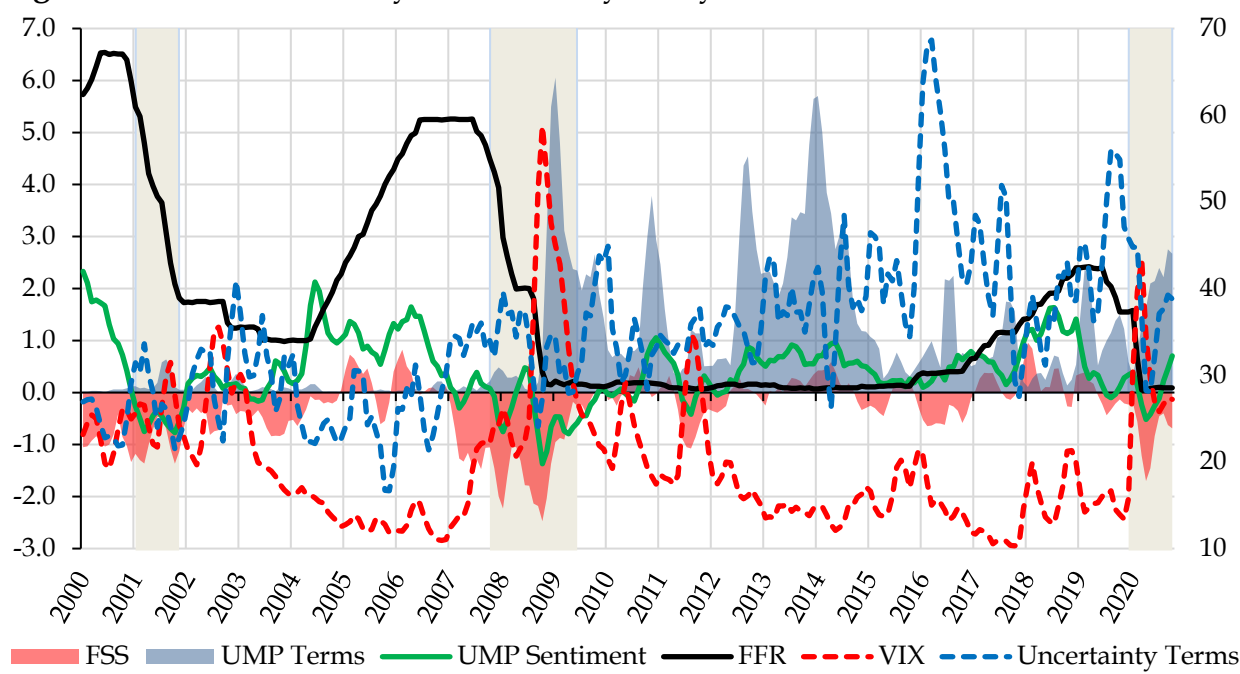

*Notes*: The gray shaded areas represent NBER recession periods. The right axis indicates the VIX and uncertainty terms levels.

Figure 13 substantiates the assertion that, during crises, the Fed diminishes the sentiment associated with UMP measures in tandem with a reduction in the frequency of uncertainty-related terms in its communications. This coordinated approach may reflect a strategy to balance policy transparency with the need to maintain market confidence during periods of stress.

The pattern of declines in FSS generally anticipating increases in the VIX finds an exception in the COVID-19 crisis. This aberration may be explained by the inherently unpredictable nature of the COVID-19 crisis compared to the more foreseeable dot-com and GFC crises, which had more conventional economic origins. The pandemic's unprecedented combination of supply and demand shocks, coupled with public health dimensions, created a unique context for financial stability considerations.

A paradigm shift occurred in the aftermath of the GFC, defining a new normal for the Fed's communications. This shift was marked by an increasing reliance on discussions centered around UMP tools, particularly forward guidance measures. These UMP deliberations featured high levels of contextual uncertainty. However, the established trajectory of this new normal encountered disruptions during the COVID-19 crisis, prompting the Fed to adopt a communication strategy that entailed a reduced utilization of uncertainty-related terms in their UMP communications.

While the preceding discussions predominantly delved into matters of financial stability, it is imperative to acknowledge that the Fed's communications equally addressed concerns pertaining to economic stability. Figure 14 provides a comprehensive overview of the word-counting indicators, as previously discussed, specifically focusing on the exchange rate.

**Figure 14.** Financial Stability, NEER, and FFR Announcements

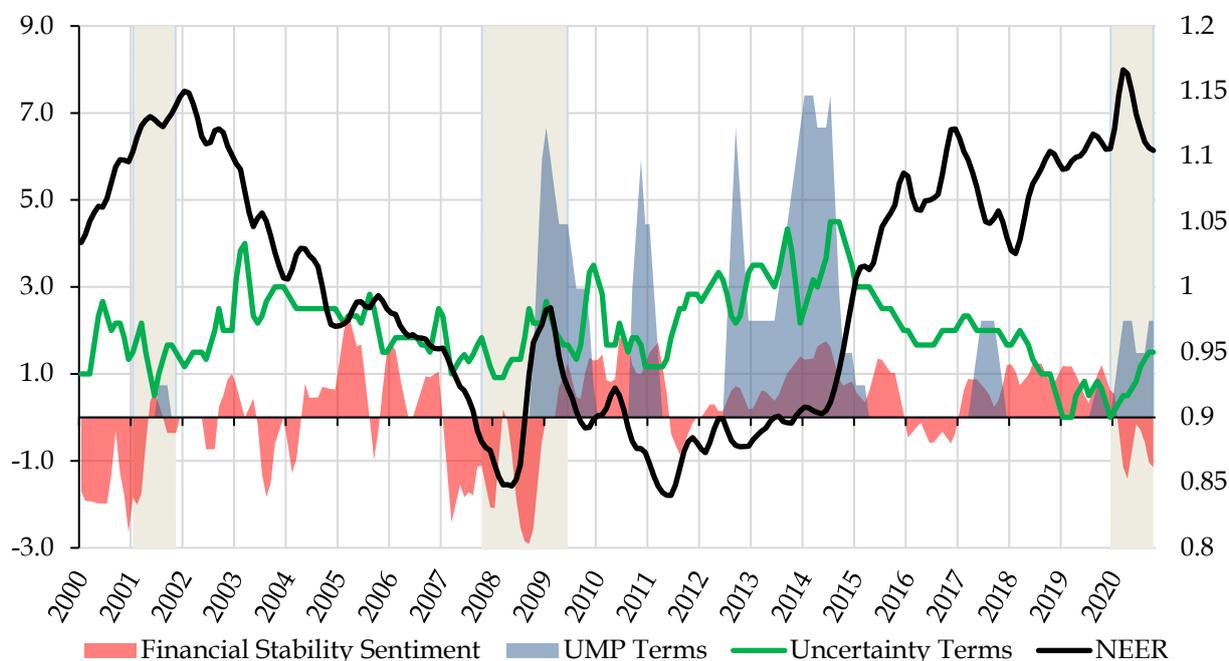

*Notes*: The gray shaded areas represent NBER recession periods. The NEER corresponds to the amount of US dollars needed to purchase foreign currency (right axis). The FSS, UMP, and uncertainty terms are related to FFR announcements.

Figure 14 illustrates a correlation between the NEER and the frequency of uncertainty-related terms within FFR announcements. The figure suggests that UMP communications and actions generally lead to a reduction in the NEER, although there is an increase during the tapering period. This relationship between UMP communications and exchange rate movements aligns with findings by Belke et al. (2017), who analyze the influence of QE on exchange rates through interest rate differentials.

Figure 15 establishes a connection between the sentiment indicators discussed earlier and another metric of economic stability: the unemployment rate. It highlights an almost consistently inverse relationship between sentiment and the unemployment rate. Positive aggregate sentiment tends to coincide with a decline in the unemployment rate, while a shift from positive to negative sentiment typically aligns with an increase in unemployment. This relationship suggests that Fed communication sentiment may reflect or anticipate labor market conditions, though we caution against inferring causality given the complex dynamics involved, especially during the COVID-19 crisis.

**Figure 15.** Sentiment and Unemployment

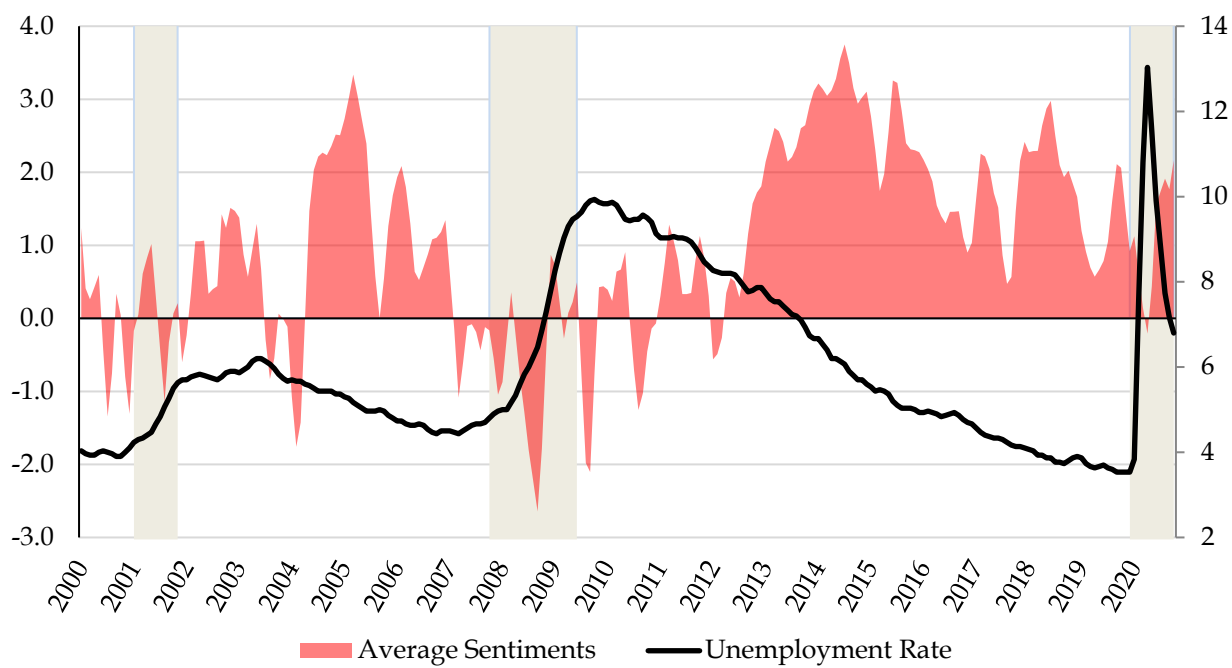

*Notes*: The gray shaded areas represent NBER recession periods. The right axis indicates the unemployment rate levels. The average sentiment aggregate contains an equally weighted average of sentiments according to LM (score and polarity), Hu&Liu (polarity), Jockers (polarity), NRC (polarity), SentiWords (polarity), UMP (score), and financial stability (score) sentiments.[20]

In contrast to previous crises, the COVID-19 crisis began with a rapid surge in the US unemployment rate, soaring from 3.5% to 13% within a brief period (January to May 2020). Subsequently, the unemployment rate declined to lower levels after May 2020, settling around 6%. The abrupt and short-lived shock to aggregate sentiment was even briefer than the corresponding impact on the unemployment rate, suggesting a crisis-specific communication strategy geared towards conveying optimism during crises to support economic confidence.

---

[20] To achieve a balanced aggregated indicator for each communication type, we weigh this average sentiment aggregate for FFR announcements more than for FOMC minutes, which in turn are weighted more than for Chairman speeches.

Figure 15 reaffirms the observation that the Fed adopted a communication strategy aimed at promoting positive sentiments during the COVID-19 crisis, potentially to counterbalance the extraordinary economic disruption caused by the pandemic and support confidence in the recovery process.

In summary, we demonstrate that the FSS is communicated through FFR announcements and FOMC minutes but is not as significantly expressed in speeches. While rises in FSS typically herald increases in the FFR, this pattern is disrupted when UMP terms are introduced into the Fed's communications, often indicating actions taken to enhance financial stability.

In addition, we have shown that the positive aggregated sentiment within the Fed's primary communications aligns with a decrease in unemployment. With few exceptions, this relationship holds during periods of substantial UMP measures. Concurrently, the NEER correlates with the degree of contextual uncertainty in the Fed's communications, highlighting the multifaceted relationships between communication patterns, economic outcomes, and financial market variables.

## 7. Policy Implications

The Fed executed a more extensive array of unconventional monetary policies during the COVID-19 pandemic than it did during the dot-com and GFC episodes. This rapid deployment within a concise timeframe was necessitated by the sudden surge in adverse economic shocks triggered by COVID-19 restrictions. The Fed's proficiency in crisis-specific communication and UMP tools, acquired during the dot-com and GFC crises, likely played a role in comprehending and addressing the communication challenges posed by the COVID-19 crisis. The success of these UMP measures hinged on clear, transparent communications, and active engagement with financial markets and the public.

Our study, employing both supervised and unsupervised learning methods, illustrates that the Fed's communications during the COVID-19 crisis significantly deviated from those of preceding crises in terms of conveyed sentiments and emphasized topics. Comparative analysis of the terms, sentiments, and topics within the Fed's communications related to COVID-19 and financial data confirms the adoption of a distinct communication strategy during the COVID-19 crisis, differing from strategies employed during the GFC and dot-com crises. We posit that these refined communication tactics may enhance the Fed's crisis management capabilities and contribute to more effective policy implementation during periods of economic stress.

Our analysis identifies this communication policy as one that instilled optimism in the public during the height of the pandemic while addressing (and implementing) unconventional monetary policies earlier in the crisis than observed in previous instances (Figure 5, Panel D). The significance of these policies in mitigating risks and uncertainties is suggested by Figures 10 and 11. Another critical observation underscores the Fed's forward-looking narrative and its adept use of communication to convey a determined sentiment and rationalize UMP measures preceding each wave of the virus or worsening financial conditions due to pandemic spillovers.

In summary, our findings demonstrate that communications explaining the adopted policies and emergency programs succeeded in portraying them as effective tools supporting economic recovery. The Fed employed a distinct communication strategy tailored to the COVID-19 crisis, projecting reduced uncertainty and less negative sentiment to the public while advocating for UMP measures to manage the crisis.[21] This strategic approach aimed at conveying optimism without compromising transparency. The Fed's timely communications, coupled with decisive actions, effectively contributed to stabilizing financial markets in conjunction with UMP and fiscal policy.

There are several potential reasons behind the Fed's communication shifts over time. Ball (1994) highlights the tension between transparency and time inconsistency, suggesting that the Fed's enhanced clarity might aim to strengthen its commitment to long-term goals. Additionally, Bernanke et al. (1999) emphasize the role of managing expectations, implying that the Fed's communication could target anchoring inflation expectations for greater policy effectiveness. Furthermore, Woodford (2003) underscores the importance of central bank credibility, suggesting that the Fed's transparency might seek to bolster public trust in its policy decisions.

Moving beyond established theories, careful interpretation of the specific changes implemented—such as the increased frequency of economic projections or the adoption of "dot plots"—could reveal nuanced motivations. For instance, employing dot plots, as analyzed by Wright (2012), might aim to enhance market understanding of individual policymakers' views and foster consensus building within the FOMC. Finally, acknowledging the limitations of any single interpretation, future research, as advocated by Orphanides and Williams (2005), could employ econometric techniques to rigorously evaluate the impact of the Fed's communication changes on market expectations and economic outcomes, providing further clarity on the central bank's communication motives and their effectiveness.

## 8. Conclusion

This study provides a comprehensive examination of central bank communication over the past two decades, with a particular focus on the COVID-19 pandemic. Our analysis reveals distinct features in Fed communications during the COVID-19 crisis compared to prior crises, including the GFC and the dot-com bubble. By correlating essential communication terms, sentiments, and topics conveyed by the Fed to COVID-19 case numbers and financial data, we distinguish the adoption of a specific communication strategy during the COVID-19 crisis.

---

[21] The significant Granger causality from UMP terms to Fed assets during COVID-19 across all communication types (F = 10.54-17.82, p < 0.01) quantifies this distinct strategy.

Throughout the COVID-19 pandemic, Fed speeches highlighted topics related to social welfare, diverging from the emphasis on policy interventions seen in FFR announcements and minutes. Discussions around COVID-19 and UMP often delved into market volatility, uncertainty, and financial stability. The sentiment of Fed communication exhibited significant changes during the COVID-19 crisis compared to the post-GFC period. Post-GFC, discussions on UMP became a "new normal" in the Fed's minutes and Chairman speeches. Additionally, we demonstrate that a negative FSS typically precedes conventional monetary policy accommodation, with exceptions during the ZLB period when conventional policy tools were constrained.

The COVID-19 crisis induced structural shifts in the content of Fed communication, suggesting the implementation of a distinct communication policy compared to the dot-com and GFC crises. This adaptive approach to crisis communication reflects the Fed's institutional learning and its evolving understanding of how communication can enhance policy effectiveness during periods of economic stress.

Our research contributes to the growing literature on central bank communication strategies (Hansen et al., 2018; Gardner et al., 2022; Bennani and Romelli, 2024) by documenting specific changes in communication content, sentiment, and timing during different crisis periods. The findings suggest that central banks can adapt their communication approaches to the unique characteristics of different economic crises, potentially enhancing policy effectiveness.

Future research avenues may explore the impact of these communication indicators on changes in interest rates, market-based expectations, exchange rates, and various asset prices within short time intervals (e.g., 30 minutes) surrounding policy announcements, building on the work of Gürkaynak et al. (2020). Such analysis could provide a more precise identification of the causal effects of specific communication strategies on financial markets and economic expectations. Additionally, extending this analysis to other major central banks would enable comparative studies of communication strategies across different institutional contexts and economic environments.